\documentclass[iop]{emulateapj-rtx4}

\shorttitle{An AGN in the NGC~3256 Southern Nucleus}
\shortauthors{Ohyama, Terashima, \& Sakamoto}

\begin{document}

\title{INFRARED AND X-RAY EVIDENCE OF AN AGN IN THE NGC~3256 SOUTHERN NUCLEUS}

\author{Youichi Ohyama\altaffilmark{1}, Yuichi Terashima\altaffilmark{2}, and Kazushi Sakamoto\altaffilmark{1}}
\affil{$^1$Academia Sinica, Institute of Astronomy and Astrophysics, No.1, Sec. 4, Roosevelt Rd, Taipei 10617, Taiwan, R.O.C.}
\affil{$^2$Department of Physics, Ehime University, Matsuyama, Ehime 790-8577, Japan}

\begin{abstract}
We investigate signs of Active Galactic Nucleus (AGN) in the luminous infrared galaxy NGC~3256 at both infrared and X-ray wavelengths.
NGC~3256 has double, the Northern and Southern, nuclei (hereafter, N and S nuclei, respectively).
We show that the {\it Spitzer} IRAC colors extracted at the S nucleus are AGN-like, and the {\it Spitzer} IRS spectrum is bluer at $<6$~$\mu$m than at the N nucleus.
We built for the S nucleus an AGN-starburst composite model with a heavily absorbed AGN to successfully reproduce not only the IRAC and IRS specrophotometries at $\simeq 3$\arcsec~but also the very deep silicate 9.7~$\mu$m absorption observed at 0\farcs 36 scale by D{\'{\i}}az-Santos et al.
We found a 2.2$~\mu$m compact source at the S nucleus in a {\it HST} NICMOS image and identified its unresolved core (at 0\farcs 26 resolution) with the compact core in previous mid-infrared observations at comparable resolution.
The flux of the 2.2$~\mu$m core is consistent with our AGN spectral energy distribution model.
We also analyzed a deeper than ever {\it Chandra} X-ray spectrum of the unresolved (at 0\farcs 5 resolution) source at the S nucleus.
We found that a dual-component power-law model (for primary and scattered ones) fits an apparently very hard spectrum with a moderately large absorption on the primary component.
Together with a limit on equivalent width of a fluorescent Fe-K emission line at 6.4 keV, the X-ray spectrum is consistent with a typical Compton-thin Seyfert 2.
We therefore suggest that the S nucleus hosts a heavily absorbed low-luminosity AGN.
\end{abstract}


\section{INTRODUCTION\label{introduction}}

Occurrence of Active Galactic Nuclei (AGNs) in Luminous Infrared Galaxies (LIRGs) has been extensively discussed since there are many lines of evidence for direct and causal relationship between them.
LIRGs predominantly emit at far infrared (FIR), and the infrared (IR) luminosities at 8--1000~$\mu$m, $L_{\rm IR}$, amount to $11<$ log $L_{\rm IR}$ (L$_{\rm \odot}$) $<12$ by definition \citep{sanders96}.
Although bulk of the energy in LIRGs is due to intense star formation activity \citep{sanders96}, AGNs are often found among LIRGs.
Also, a significant fraction of LIRGs are tidally interacting objects.
The fraction of AGN-dominated sources is known to be higher in LIRGs with larger IR luminosity (e.g., \citealt{veilleux95,kim95,nardini08,nardini10,petric11,iwasawa11,ah12}) and in later interaction stages \citep{petric11,iwasawa11}.

Both mid-infrared (MIR) and X-ray studies are very effective to reveal nuclear activities in LIRGs.
Although LIRGs, in particular their nuclei, are often heavily absorbed at optical and near-infrared (NIR) wavelengths, both MIR and hard X-ray photons can penetrate through such interstellar medium.
We can study AGN indications at MIR because AGNs typically show characteristic power-law-like spectral energy distribution (SED), which often shows an excess at MIR over the SED from warm and cold dusts (e.g., \citealt{granato97,ah01,nenkova02,armus07,donley12}).
Various kinds of AGN diagnostics have been developed at MIR, and empirical assessments of relative AGN contribution to the entire system have been made \citep{genzel98,strum02,armus07,spoon07,desai07,petric11,stierwalt13}.
We can also study typical AGNs at X-ray because AGNs show luminous and characteristic power-law spectra, the hard part of which suffers from less absorption by circum-AGN material.
Therefore, X-ray has been heavily utilized to study AGN in LIRGs (e.g., \citealt{iwasawa11}).

NGC~3256 is the most luminous LIRG in the local universe ($z<0.01$).
At a distance of 35~Mpc (for consistency with \citealt{sakamoto14}, which adopted \citealt{sanders03}\footnote{\cite{sanders03} adopted distance with $cz$ using the cosmic attractor model outlined in Appendix A of \cite{mould00}, using $H_{\rm 0}=75$ km s$^{-1}$ Mpc$^{-1}$ and adopting a flat cosmology in which $\Omega_{\rm M}=0.3$ and $\Omega_{\rm \lambda}=0.7$.}), its IR luminosity ($L_{\rm IR}$) is as large as $3.6 \times 10^{11}$ $\rm {L}_{\rm \odot}$ \citep{sanders03}, or total IR luminosity at 3--1100~$\mu$m (TIR) is (2.7--3.3)$ \times 10^{11}$ $\rm{L}_{\rm \odot}$(\citealt{engelbracht08} converted for our assumed distance).
It is also among the most luminous X-ray sources without a confirmed AGN in the local universe \citep{moran99,lira02,ps11}.
It is a major merging system showing tidally distorted morphology at galaxy scale with tidal tails, disturbed outer spiral arms, and prominent dust lanes with complex morphology (e.g., \citealt{graham84,kotilainen96,zepf99,lipari00,ah02,english03,ah06a,ah06b,sakamoto14}).
Its double nuclei, often referred to as Northern and Southern nuclei (hereafter, N and S nuclei, respectively) are separated by only 5\arcsec~(850~pc).
They are clearly visible at both radio and MIR (e.g., \citealt{norris95,kotilainen96,neff03,ah06b,lira08,sakamoto14}), but the S nucleus is hidden in dust at the optical.
The merger is likely in the late stage just before coalescence of the nuclei \citep{goals,stierwalt13}.

The N nucleus is a core of a starburst galaxy.
Its optical spectrum shows H~{\sc ii} region-like features.
Extended outflowing ionized gas with shocks is also detected and attributed to a superwind powered by the starburst \citep{moran99,lipari00,ifu1,ifu2}.
Stellar population examined in $K$ band indicates young starburst population \citep{doyon94}.
Prominent Polycyclic Aromatic Hydrocarbon (PAH) features as well as low-ionization fine structure lines have been detected at MIR \citep{siebenmorgen04,mh06,lira08,ps10,ds10}, also indicating star formation activity.
The nucleus is spatially resolved with {\it HST} at 1.6~$\mu$m at 0\farcs 14 full-width half maximum (FWHM) resolution \citep{ah02}, the 8.1~m Gemini Telescope at 8.74~$\mu$m at its diffraction-limit (0\farcs 30 FWHM) resolution \citep{ah06b}, and {\it Chandra} at X-ray at 0\farcs 5 FWHM resolution \citep{lira02}.

Presence of an AGN at the optically obscured S nucleus has been suspected for a long time (e.g., \citealt{kotilainen96}), but there is still no firm evidence for it.
Dust lanes cover the S nucleus and make it invisible at optical and NIR wavelengths below $K$ band.
A compact (unresolved at 0\farcs 5 FWHM) and moderately absorbed (neutral Hydrogen column density $N_{\rm H} \simeq 5 \times 10^{22}$ cm$^{-2}$) X-ray source was detected with {\it Chandra} at this nucleus \citep{lira02}.
Although it looks much fainter (by at least two orders of magnitude) than expected for a classical Seyfert nucleus given its MIR luminosity, it is brighter than expected for typical starburst galaxies (\citealt{lira02}; see also \citealt{awaki91,turner97,guainazzi05,fukazawa11}).
Analysis of the MIR SED extracted within a kpc-scale aperture (including both nuclei and the host galaxy component) indicated that AGN contribution to the total MIR luminosity is $<5$\% \citep{ah12}.
\cite{sakamoto14} recently discovered a bipolar collimated jet-like molecular-gas outflow from the S nucleus, and argued that it is likely driven by an AGN.

Due to its proximity, NGC~3256 is among the best targets to examine the characteristics and roles of AGN, if any, within a LIRG system.
For this purpose, it is essential to verify the presence of an AGN in the S nucleus using spatially resolved and/or sensitive MIR and X-ray observations (e.g., \citealt{lira02,ds10b,ds11,ps10}).
Recent spatially-resolved MIR spectroscopies revealed some interesting differences between the two nuclei.
By using the slit-scanning spectroscopy data taken with the {\it Spitzer} IRS \citep{irs}, \cite{ps10} noted that the S nucleus shows deeper silicate absorption at 9.7~$\mu$m, stronger H$_2$ lines, and a slightly smaller equivalent width (EW) of PAH 6.2~$\mu$m than those at the N nucleus.
More dramatic differences are found in ground-based spectra obtained with higher-spatial resolution.
The silicate absorption is much deeper and the PAH features are undetectably weaker at 0\farcs 36 (61~pc) scale at the S nucleus than in the IRS spectrum \citep{ds10}.
At X-ray, \cite{lira02} reported that the point-like source at the S nucleus is more heavily absorbed than other point sources and the diffuse component within the galaxy.
In this work we utilize archival and published data of {\it Spitzer}, {\it HST}, and {\it Chandra} to analyze the nuclear spectrophotometry more comprehensively than ever to explore an AGN in the S nucleus.

\section{INFRARED DATA}

\subsection{{\it Spitzer} IRAC\label{irac_data_analysis}}

\begin{center}
\begin{figure}
\includegraphics[scale=0.8]{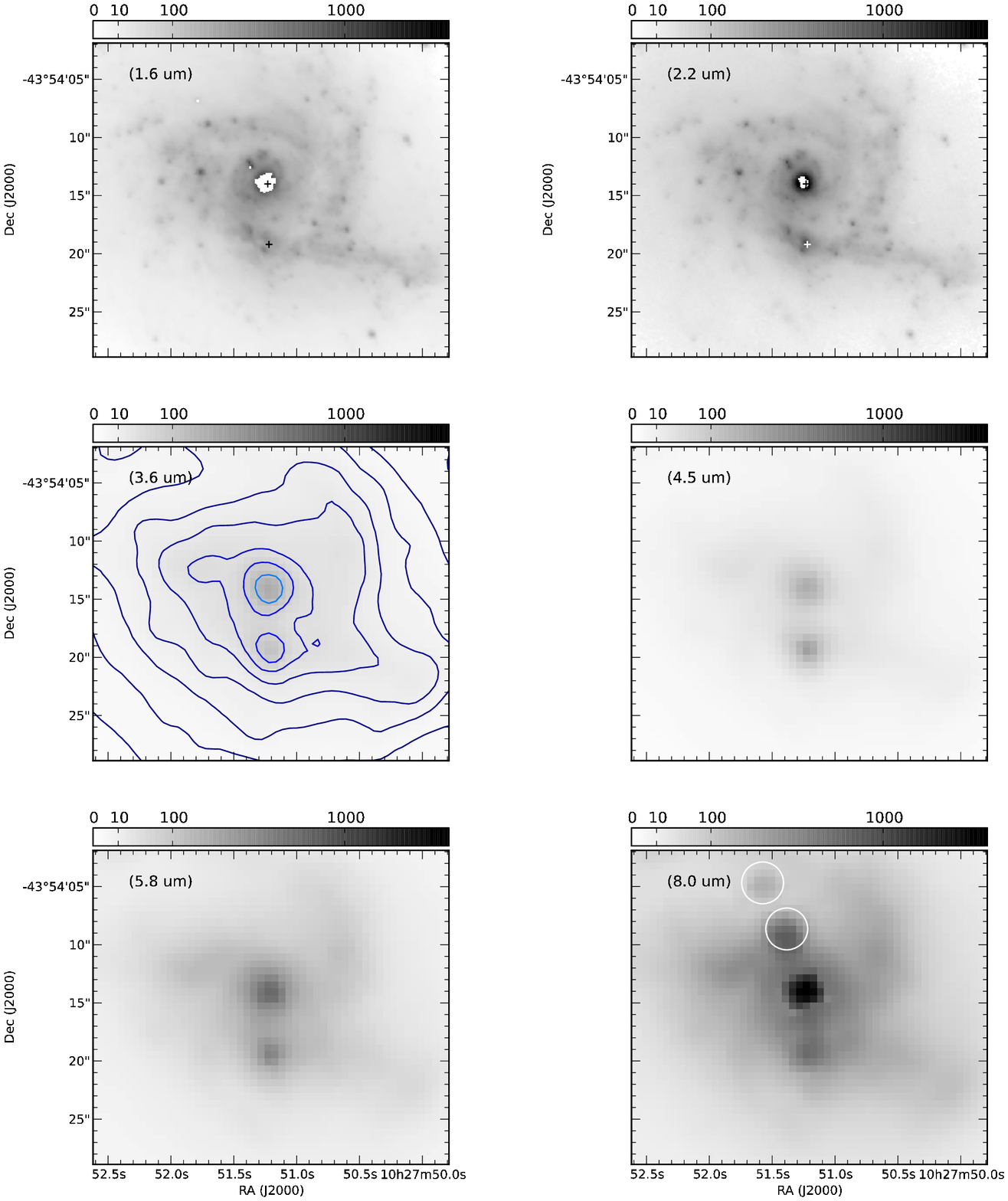}
\caption{
{\it HST} NICMOS F160W and F222M images (upper row) and {\it Spitzer} IRAC four channel images (middle and lower rows) of NGC~3256 around the N and S nuclei.
North is up and east is to the left.
Both NICMOS and IRAC images are shown in a square root scale as indicated in scale bars (in units of MJy/steradian) for each panel.
The N nucleus is saturated in the two NICMOS images, and are masked.
Positions of the two nuclei in radio continuum \citep{neff03} are marked with crosses on the NICMOS images.
Contours at logarithmic steps are overlaid on the IRAC 3.6~$\mu$m image.
In the lower right panel (IRAC 8.0~$\mu$m), two white circles indicate locations of artifacts (known as the ``bandwidth effect''; IRAC instrument handbook) caused by the saturated N nucleus.
Also in this panel, the saturated pixels around the N nucleus are fixed as described in the main text.
\label{fig_morphology}}
\end{figure}
\end{center}

We examined {\it Spitzer} IRAC \citep{irac} archival images of NGC~3256 at its four channels for morphological and photometric studies.
Although \cite{lira08} (see also \citealt{lira02}) published the IRAC images and provided nuclear photometries, they did not use the 8.0~$\mu$m channel image in their analysis because it is slightly saturated at around the N nucleus.
The 8.0~$\mu$m information is helpful in conjunction with information of the 3.6, 4.5, and 5.8~$\mu$m channels to create IRAC color-color diagrams for AGN diagnostics (e.g., \citealt{lacy04,sajina05,stern05,donley12}).
Fortunately, the images were taken under the HDR (High Dynamic Range) mode in which both short- and long-exposure frames are taken within an Astronomical Observation Request (AOR).
The short-exposure frame is not saturated at the N nucleus.
We retrieved the Post Basic-Calibrated Data (Post BCD or PBCD) of standard mapping observations (AOR: 3896832; PI: G. Fazio) from the {\it Spitzer} Heritage Archive.
This is the same data set of \cite{lira02,lira08}.
For the 8.0~$\mu$m channel, since short-exposure frames are not recommended for science use in general (IRAC data handling book), we confirmed flux calibration consistency between the long- and short-exposure frames using photometry of field stars.
Then the saturated pixels at around the N nucleus in the long-exposure frame were substituted with corresponding pixels in the short-exposure frame image.
For the 4.5~$\mu$m channel, additional simple sky subtraction was made to correct for its tilted background.
As shown by \cite{lira08} for the 3.6, 4.5, and 5.8~$\mu$m channels, the two compact nuclei are clearly separated in all four channels (Figure~\ref{fig_morphology}).

\begin{center}
\begin{figure}
\includegraphics[scale=0.6,angle=90]{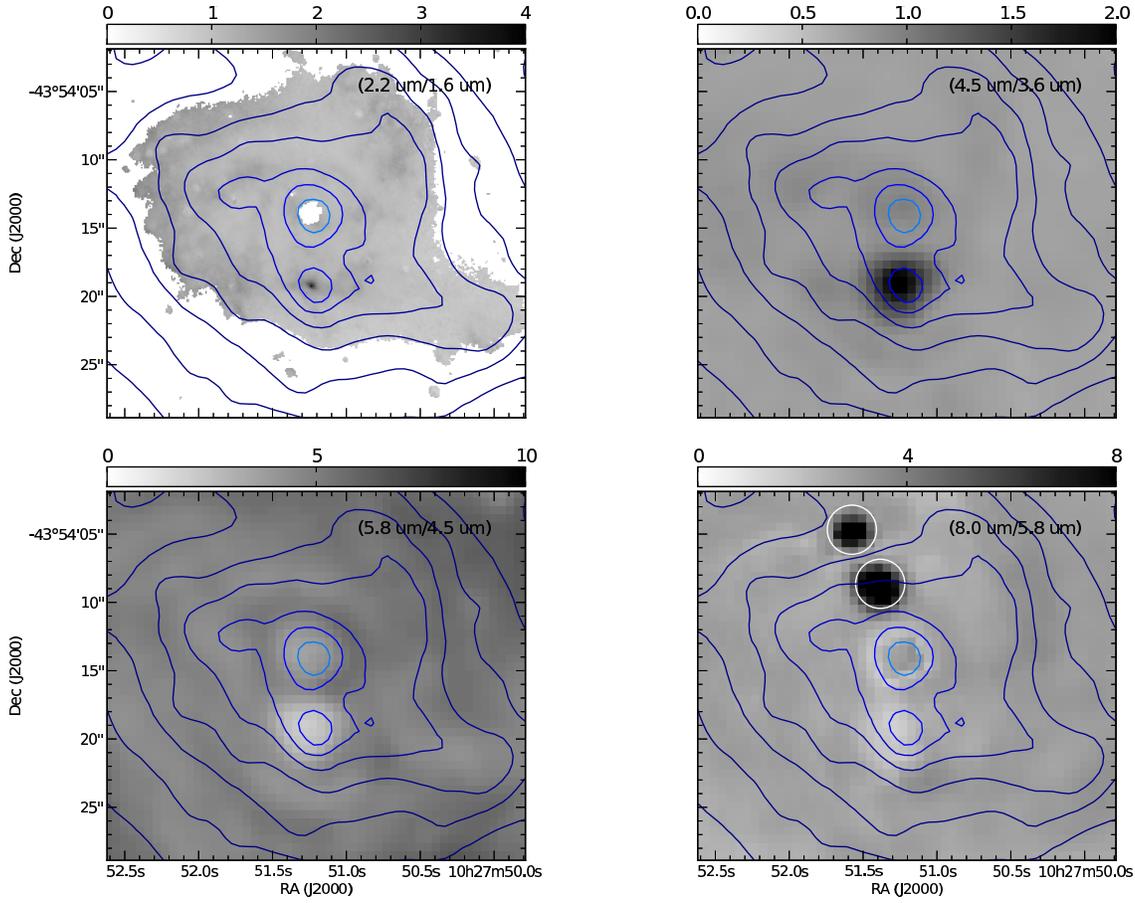}
\caption{
Flux ratio maps of NICMOS F222M/F160W (upper left), IRAC 4.5~$\mu$m/3.6~$\mu$m (upper right), 5.8~$\mu$m/4.5~$\mu$m (lower left), and 8.0~$\mu$m/5.8~$\mu$m (lower right) around the N and S nuclei.
All images are shown in the same sky area as in Figure~\ref{fig_morphology}.
Contours of the IRAC 3.6~$\mu$m image shown in Figure~\ref{fig_morphology} are overlaid on all flux ratio maps for positional references.
Flux ratio scales are indicated in scale bars for each panel.
Outer fainter regions with signal-to-noise ratio being less than 3.0 per pixel are masked in the F222M/F160W image.
Two white circles in the 8.0~$\mu$m/5.8~$\mu$m map (lower right) indicate areas that are affected by the artifacts in the 8.0~$\mu$m image as shown in Figure~\ref{fig_morphology}.\label{fig_ratioimage}}
\end{figure}
\end{center}

Aperture photometry was made on each nucleus with our IDL program (Table~\ref{tab_irac_flux}).
We used an aperture of 2\farcs 8 diameter for the photometry, and applied aperture correction following \cite{apcor}.
Flux contribution from the host galaxy is estimated within a concentric annulus whose inner and outer radii are 2\farcs 0 and 2\farcs 6, respectively.
The size of the annulus was set as small as possible to estimate the flux level near the nucleus by minimizing the effect of host galaxy structure.
Standard deviation within the annulus is quadratically added to the statistical error of the flux measurement of the nuclei.
Although our fluxes are systematically slightly fainter than those of \cite{lira08}, the two are consistent because \cite{lira08} performed simple aperture photometry with slightly larger apertures than ours (3\farcs 6, 3\farcs 6, and 4\farcs 0 diameter apertures for the 3.6~$\mu$m, 4.5~$\mu$m, and 5.8~$\mu$m images, respectively).
We note that N nucleus has only about 7\%, 7\%, 6\%, and 7\%, and the S nucleus has about 4\%, 9\%, 4\%, and 2\% in 3.6~$\mu$m, 4.5~$\mu$m, 5.8~$\mu$m, and 8.0~$\mu$m channels, respectively, of the flux integrated over the entire galaxy reported by \cite{engelbracht08}.
Flux ratio maps were also made for 4.5~$\mu$m/3.6~$\mu$m, 5.8~$\mu$m/4.5~$\mu$m, and 8.0~$\mu$m/5.8~$\mu$m (Figure~\ref{fig_ratioimage}).
The point spread function (PSF) size matching was made for these ratio maps using a bright field star as a reference.
The measured PSF FWHM is about 2.9 pixel or 1\farcs 7.

\subsection{{\it Spitzer} IRS\label{irs_data_analysis}}

We processed our own IRS spectra so that the spectrophotometry was done in a way consistent with the corresponding IRAC photometry for each nucleus, although similar spectra had been already published \citep{brandl06,bs09,ps10,ds10,ah12}.
We used the same IRS mapping data set of \cite{ps10} and \cite{ah12}, and followed almost the same processing methods as adopted in the earlier studies.
The BCD data (AOR: 17659904; PI: G. Rieke) of the Short-Low (SL) module were retrieved from the {\it Spitzer} Heritage Archive.
Similar mapping data with other modules (Long-Low (LL), Long-High (LH), and Short-High (SH)) were not used since spatial resolution was worse for the longer wavelength data.
They are not suitable for our analysis in which separating the two nuclei is essential.
The standard CUBISM data processing \citep{cubism,ps10} was made, including background subtraction, hot/rogue pixel identification and subtraction, and cube creation.

Aperture photometry was made with our IDL program to extract spectra for each nucleus on the data cubes created with CUBISM (Figure~\ref{fig_irac_irs_sed}).
We fixed an aperture size of 3\farcs 6 diameter over 5.2--14.5~$\mu$m and applied the aperture correction.
Flux contribution from the host galaxy is estimated within a concentric annulus whose inner and outer radii are 2\farcs 0 and 2\farcs 6, respectively.
The size of the annulus was set as small as possible to estimate flux level near the nucleus by minimizing the effect of host galaxy structure.
However, since the IRS spatial resolution ($\simeq 4$\arcsec~FWHM) is larger than that of IRAC and is comparable to the distance to another nucleus ($\simeq $5\arcsec), we carefully applied masks to hide another nuclear component in the annulus.
Standard deviation within the annulus is quadratically added to the statistical error of the flux measurement of the nuclei.
The aperture correction factor was measured with a mapping observation of a standard star, HR~7341.
The standard star data (AOR 19324160) taken in the same observing campaign of the NGC~3256 observation were retrieved from the {\it Spitzer} Heritage Archive, and they were reduced by CUBISM with exactly the same parameters.
We then measured the star with the same photometry program and aperture settings, and found the aperture correction factor by comparing the measured spectrum with the standard-star flux information from the IRS Instrument Handbook.

As in the case of the IRAC photometry, those nuclear spectrophotometries are about 12--20\% of that over the entire galaxy (13\farcs 4 $\times$ 13 \farcs 4; \citealt{ah12}), although the overall spectral shapes look similar, i.e., all show prominent PAH 6.2, 7.7, 8.6, 11.3~$\mu$m features and fine structure lines such as [Ne~{\sc ii}].
We confirmed characteristics reported earlier by \cite{ps10}, namely brighter H$_2$ fluxes, deeper silicate absorption at 9.7~$\mu$m, and smaller EW of PAH 6.2~$\mu$m at the S nucleus (see \S \ref{infrared_agn_signature_results} below for our silicate absorption depth measurements).

\subsubsection{Flux Matching between the IRAC and IRS Spectrophotometries}\label{irac_iras_fluxmatch}

The IRS nuclear photometries are about two times larger than the IRAC 5.8~$\mu$m and 8.0~$\mu$m nuclear photometries at both nuclei (Figure~\ref{fig_irac_irs_sed}), and we interpret this as a result of more contamination from the host component due to larger PSF of IRS.
Therefore, in order to combine the IRS and IRAC spectrophotometries in a consistent way for our subsequent analyses, we adjusted the IRAC nuclear photometries for the IRS resolution in the following way.
For the N nucleus, we simply scaled the observed nuclear IRAC fluxes by a factor measured at both 5.8~$\mu$m and 8.0~$\mu$m channels by comparing with synthetic IRAC fluxes from the IRS spectrum.
We utilized a standard photometry tool (spitzer\_synthphot.pro) from {\it Spitzer} Data Analysis Cookbook.
Since the IRS SL module cannot fully cover the IRAC 5.8~$\mu$m bandpass between 4.9 and 5.2~$\mu$m, we extrapolated the spectra shortward a little with a help of scaled SWIRE NGC~6090 SED template (see \S \ref{infrared_agn_signature_results} below for more details about the template).
We found that a single scaling factor on the IRAC N nuclear fluxes satisfies matching both 5.8~$\mu$m and 8.0~$\mu$m fluxes with the IRS-synthetic photometries, and the factor is applied to all four IRAC channel fluxes.
For the S nucleus, we added scaled IRAC fluxes at the N nucleus to the fluxes at the S nucleus to match IRS-synthetic photometries at the S nucleus.
Here we utilized the fact that colors of the N nucleus are very similar to those in the surrounding region of the S nucleus (Figure~\ref{fig_ratioimage}).
We calculated the scaling factor by considering, again, both 5.8~$\mu$m and 8.0~$\mu$m fluxes.
We found that a single scaling factor satisfies matching fluxes in both channels, and applied the factor to all four IRAC channels.
The adjusted IRAC fluxes for each nucleus are also shown in Table~\ref{tab_irac_flux}.
Note that the adjusted fluxes are used only for analyzing MIR SED together with IRS.

\subsection{T-ReCS Spectrum\label{trecs_data_analysis}}

The $N$-band high-spatial low-spectral resolution spectrophotometry of the S nucleus taken by \cite{ds10} shows remarkably different signatures just on the nucleus, and we therefore include the data in our analysis.
The spectrum was taken with T-ReCS at the Gemini-South Telescope through a 0\farcs 36-wide slit under 0\farcs 35 FWHM seeing condition, providing the best spatial resolution at $N$ band owing to both the diffraction-limited 8~m telescope and the seeing-matched narrow slit.
The spectrum is extracted over 0\farcs 36 (or four pixel width) around the peak at the S nucleus.
\cite{ds10} performed flux calibration by a standard star observation while considering the slit loss effect, and we further applied the aperture correction to estimate the flux of the central compact source by compensating loss of photons outside the extraction aperture box.
By assuming a Gaussian PSF of the seeing size, we estimated and multiplied a factor of 1.30 on the original data presented by \cite{ds10} in their Figure~3.
Since error of the spectrum is not explicitly given, we estimated it from a scatter of the data points around the fitted model profile (see Figure~3 of \citealt{ds10}).

\subsection{{\it HST} NICMOS Images\label{nicmos_data_analysis}}

In order to extend the shorter wavelength coverage of the nuclear spectrophotometries, we also analyzed high-resolution {\it HST} NICMOS images at NIR.
We retrieved the combined NIC3 images of F110W (1.1~$\mu$m wide band), F160W (1.6~$\mu$m wide band), and F222M (2.2~$\mu$m medium-wide band) from the {\it Hubble} Legacy Archive.
The NIC3 camera provides a larger field-of-view and is good for comparison with the IRAC images.
Their spatial resolutions (PSF size in FWHM) are 0\farcs 26 for all three bands.
Flux measurement is based on standard flux conversion factors for the instrument and the filter from the NICMOS Handbook.
The {\it HST} images presented by \cite{lira02} and \cite{ah02,ah06a,ah06b} are from a different data set taken with the NIC2 camera of the NICMOS instrument.
A small correction of tilted background was made for each filter image by fitting the image with a tilted plane while masking sources for each filter band.
Astrometry is corrected with three bright field stars in the 2MASS catalog outside of the galaxy, and the global accuracy is about 0\farcs 2.
The F160W and F222M images are shown along with the IRAC images in Figure~\ref{fig_morphology}.
Zoom-up images of all three filters around the S nucleus are also shown in Figure~\ref{fig_nicmos_snuc}.
A compact source marginally appears on the F222M image, although it is blended with surrounding structures.
This source is not clearly identified in F160W and F110W.
For more NICMOS images for other lines/bands and their descriptions, see \cite{lira02} and \cite{ah02,ah06a}.

\section{INFRARED RESULTS}

\subsection{Infrared Morphology}\label{infrared_morphology}

Figure~\ref{fig_morphology} compares overall morphology of the galaxy in all IRAC channels and NICMOS F160W and F222M images.
In general, IRAC channels (in particular the longer 5.8 and 8.0~$\mu$m channels) trace star-forming regions and the NICMOS bands trace stellar population.
The N nucleus is bright and distinct in all IRAC channels.
The S nucleus is barely visible at 2.2~$\mu$m and is more evident in the IRAC channels.
The N nucleus is brighter than the S nucleus at 3.6, 5.8, and 8.0~$\mu$m, while the two nuclei are comparable at 4.5~$\mu$m.
Such characteristics were noted for 2.2, 3.6, 4.5, and 5.8~$\mu$m by \cite{lira02,lira08}.
\cite{kotilainen96} also showed that both nuclei are equally bright at $L$' band (3.6~$\mu$m).

The flux ratios at the S nucleus are remarkably different from the rest due to a compact red source at the S nucleus.
The S nucleus shows larger 2.2~$\mu$m/1.6~$\mu$m and 4.5~$\mu$m/3.6~$\mu$m, and smaller 5.8~$\mu$m/4.5~$\mu$m flux ratios than those of the N nucleus and most of the host galaxy (Figure~\ref{fig_ratioimage}).
The large-scale dust lanes known around the S nucleus in the optical wavelengths \citep{zepf99,moran99,lipari00,ah02,ah06a,sakamoto14} are not visible in the MIR flux ratio maps.
The red S nucleus stands out even more in the {\it HST} data when the stellar component with complex morphology is subtracted.
We modeled the stellar component of the host galaxy in the F222M band using F160W and F110W images.
First we measured stellar colors at each pixel by using the two images, and we then synthesized the stellar F222M image by linearly extrapolating from the F160W image by considering the stellar colors.
We scaled the synthesized image by a factor of 0.74 to match the observed F222M image at well outside of the S nucleus.
This correction factor is probably due to color corrections on the filters for very red stellar continuum along with the heavy dust lanes around the S nucleus \citep{zepf99,moran99,lipari00,ah02,ah06a,sakamoto14}.
We found a distinct compact peak at the S nucleus on the F222M image subtracted for its synthetic stellar component (Figure~\ref{fig_nicmos_snuc}), indicating an additional component at the S nucleus.
This compact source coincides with the red peak in the F222M/F160W image, a radio continuum source at the S nucleus (e.g., \citealt{norris95,neff03,sakamoto14}), and a compact (at 0\farcs 30 FWHM resolution) 8.74~$\mu$m continuum source at the S nucleus \citep{ah06b}.

The red compact source at the S nucleus contains an unresolved (at 0\farcs 26 FWHM, or 44 pc) core.
We fitted the source with one and two circular Gaussian components.
In the single-Gaussian model, we found a systematic radial variation in the residual map in a sense that the center has a compact positive residual peak and the surrounding region systematically shows negative residuals.
In the double-Gaussian model, we assumed that the two Gaussian components share the same center position and that the more compact component is unresolved (at 0\farcs 26 FWHM).
The residual map has much less systematic structure compared with the single-Gaussian model.
The reduced $\chi^2$ are 1.8 and 1.0 for the single- and double-Gaussian models, respectively.
Therefore, the double-Gaussian model with an unresolved core is preferred.
The fitting results are summarized in Table~\ref{tab_nicmos_measurement}.

\subsection{MIR Spectra/SEDs and Color-Based AGN Diagnostics at Individual Nuclei}\label{infrared_agn_signature_results}

\begin{figure}
\plotone{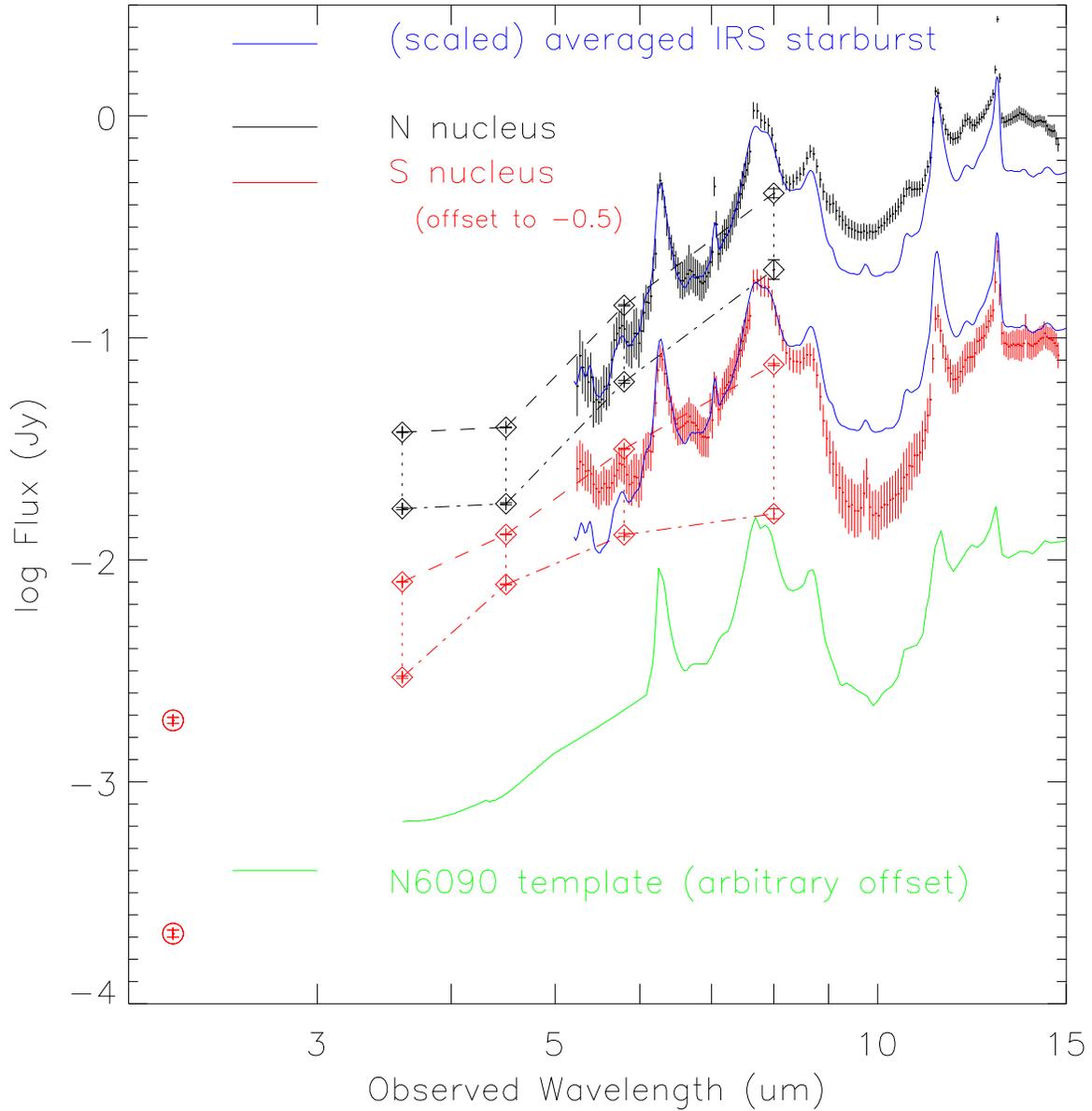}
\caption{
The IRS spectra and IRAC photometries of the N (black; top) and S (red; middle) nuclei, and the NICMOS photometry of the S nucleus (red).
All photometries for the S nucleus are shown with an offset of $-0.5$ for clarity of the figure.
IRAC fluxes of each nucleus are shown both for nuclear fluxes (diamonds connected by dot-dashed lines) and adjusted fluxes (to match the IRS spectra; diamonds connected by dashed lines.
See the main text for the details of the adjustment.
We also show the NICMOS F222M nuclear photometry of an unresolved compact source (lower) and sum of the unresolved and resolved sources (upper) at the S nucleus with red circles.
The error bars are for $\pm 1\sigma$.
Each blue solid line shows the starburst IRS spectrum of \cite{brandl06} that is redshifted to NGC~3256 and scaled at 6.0--8.0~$\mu$m for each nucleus.
The LIRG SED template of NGC~6090 arbitrary scaled for clarify of the figure is also shown with green solid line (bottom).
\label{fig_irac_irs_sed}}
\end{figure}

\begin{center}
\begin{figure}
\includegraphics[scale=0.6,angle=90]{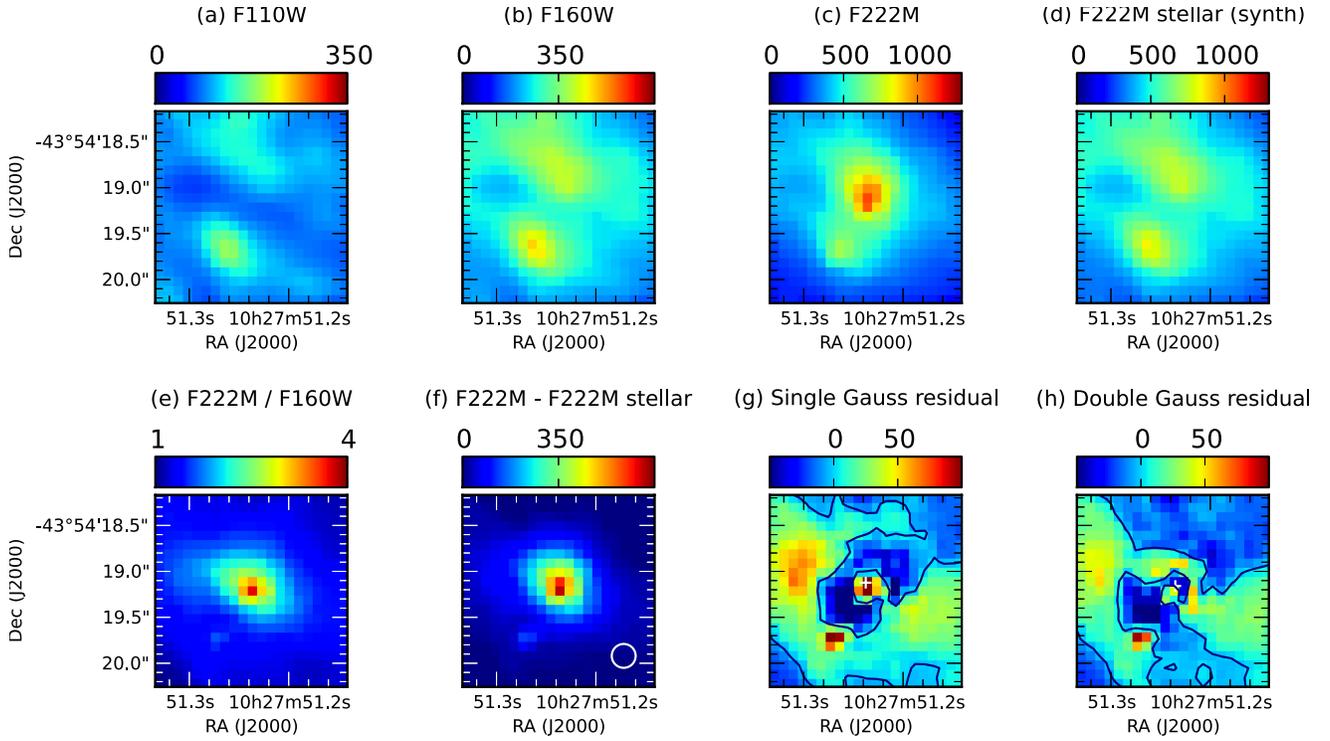}
\caption{
Zoom-up NICMOS images of the S nucleus and results of our analysis.
(a)--(c) NICMOS F110W, F160W, and F222M images.
(d) Synthetic F222M image of the stellar component.
See the main text for the details of the synthesis.
(e) A flux ratio image of F222M/F160W.
(f) The F222M nuclear component, made by subtracting the synthetic F222M stellar component image from the observed F222M image.
The NICMOS resolution (0\farcs 26 FWHM) is indicated by a white circle.
(g)--(h) The residual images of the F222M nuclear component after subtracting the fitted Gaussian models.
The panel (g) shows a case of single-Gaussian model, and (h) shows a case of double-Gaussian model.
The fitted positions of Gaussian components are marked with white pluses in panels (g) and (h).
All panels are shown in linear scale, and flux (in units of MJy/steradian) and flux ratio (for only panel f) scale are indicated in scale bars for each panel.
The zero flux levels in the residual images are also indicated by black contours.
\label{fig_nicmos_snuc}
}
\end{figure}
\end{center}

\begin{figure}
\plotone{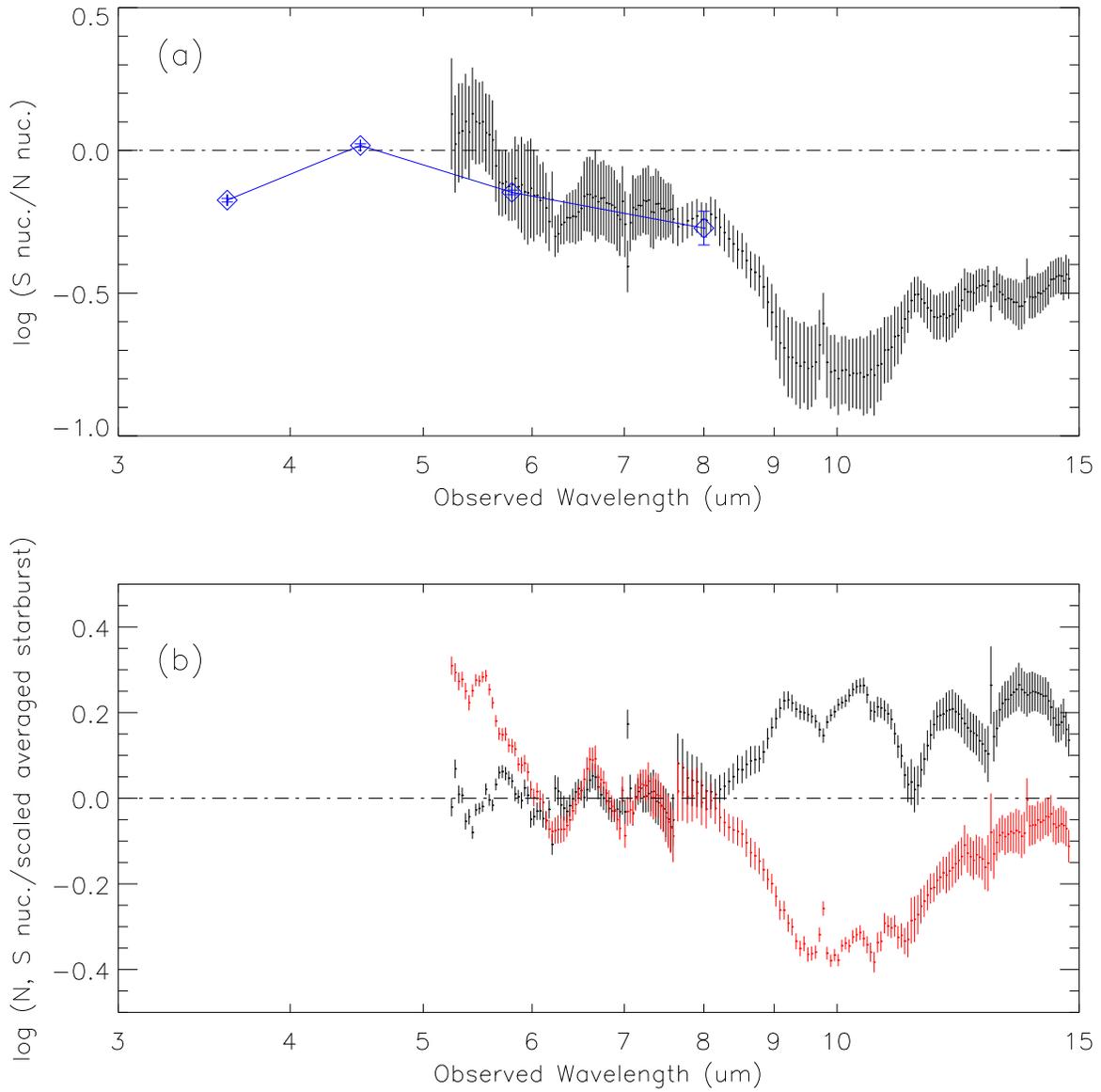}
\caption{
(a) The spectrum of the flux ratio between the N and S nuclei.
The IRS data are in black and the IRAC data are in blue.
(b) The spectra of the flux ratio between the individual nuclei and the IRS starburst template of \cite{brandl06}.
The latter is redshifted and scaled at 6.0--8.0~$\mu$m for each nucleus.
The N nucleus is shown in black and the S nucleus is in red.
Error bars are $\pm 1\sigma$.
\label{fig_irs_irac_ratio}}
\end{figure}

The two nuclei are basically similar to each other in the IRS spectra, but we found their noticeable differences.
The IRS spectra of both nuclei show prominent PAH and mild silicate absorption features as well as some fine structure lines (Figure~\ref{fig_irac_irs_sed}).
Such characteristics are common to LIRGs (e.g., \citealt{petric11}).
We compare the observations with the starburst IRS template of \cite{brandl06}, which was generated by averaging IRS spectra of local starburst galaxies.
Since this template is taken with exactly the same instrument and module (SL), we can directly compare it with our observations.
Although the template reproduces the observations rather well, we found two notable differences between them.
One is the depth of the $9.7$~$\mu$m silicate absorption, which is characterized by $S_{\rm Si~9.7~\mu m} \equiv ln~(F^{\rm obs}_{\lambda}/F^{\rm cont}_{\lambda})$.
Here, $F^{\rm cont}_{\lambda}$ is a continuum flux at 9.7~$\mu$m estimated following \cite{ds10}.
We measured $S_{\rm Si~9.7~\mu m}= -0.65\pm 0.20$, $-1.31\pm 0.34$, and $-0.50$ for the N nucleus, S nucleus, and the IRS starburst template, respectively.
The S nucleus shows significantly deeper absorption than the N nucleus, which shows similar depth within uncertainty as the starburst template.
We note that the N nucleus shows slightly smaller $S_{\rm Si~9.7~\mu m}$ than the template, although the observed 9.7~$\mu$m flux at the N nucleus is larger than the template scaled at 6.0--8.0~$\mu$m (Figure~\ref{fig_irac_irs_sed}).
This is because the N nucleus also shows higher continuum at $\gtrsim 12$~$\mu$m than the scaled template due to systematically redder color at $\gtrsim 9$~$\mu$m, and the estimated continuum level at 9.7~$\mu$m is also higher for the N nucleus.
Another difference is the color at $\lambda \lesssim 6$~$\mu$m.
Ratio of our IRS observations to the starburst template (Figure~\ref{fig_irs_irac_ratio}) shows a steep excess at the S nucleus below $\simeq 6$~$\mu$m, reaching about data-to-template ratio of $\sim 2$ at $\simeq 5$~$\mu$m.
The N nucleus does not show such a systematic deviation from the template.
We also compare the observations with a starburst-powered LIRG SED template of NGC~6090 (log $L_{\rm IR}$ ($\rm L_{\rm \odot}) =11.51$; \citealt{sanders03}) from \cite{swire} (Figure~\ref{fig_irac_irs_sed}).
The template was generated to fit the observations with the GRASIL code \citep{silva98}, which is a physical starburst evolution model for estimating SEDs.
Although this template also roughly resembles the spectra of two nuclei, it seems closer to the N nucleus at $<6$~$\mu$m.

The IRAC SEDs further illustrate the difference between the two nuclei (Figure~\ref{fig_irac_irs_sed}).
At the S nucleus, the IRAC flux monotonically increases toward longer wavelengths, i.e., no flux drop is found at 4.5~$\mu$m.
At the N nucleus, the IRAC 3.6--4.5~$\mu$m color is almost flat, indicating a contribution from the stellar component, which has a blue color at $>1.6$~$\mu$m (e.g., \citealt{sawicki02}).
The slope between 4.5~$\mu$m and 8.0~$\mu$m is steeper (redder) at the N nucleus than at the S nucleus.
An excess flux at 4.5~$\mu$m at the S nucleus causes such differences.

By utilizing IRAC color-color diagrams for AGN diagnostics, we found that the S nucleus shows AGN-like SED at 3.6--8.0~$\mu$m (Figure~\ref{fig_irac_color_color}).
Two types of such diagrams have been developed for AGN diagnostics.
One is based on flux ratios of 5.8~$\mu$m/3.6~$\mu$m vs. 8.0~$\mu$m/4.5~$\mu$m originally proposed by \cite{lacy04}.
We take color boundaries for AGN selection from \cite{donley12}, who modified the flux ratio cut of 5.8~$\mu$m/3.6~$\mu$m from the original one.
Another is based on 5.8~$\mu$m-8.0~$\mu$m and 3.6~$\mu$m-4.5~$\mu$m colors (in magnitudes) proposed by \cite{stern05}.
Although the same IRAC flux information is used in both plots, they are not completely consistent in identifying photometric AGN candidates.
The IRAC colors at the S nucleus are in the area for AGNs, which include SEDs of pure power-law and hot blackbody of 300--1000~K.
The N nucleus is closer in the IRAC colors to starburst-dominated sources (either spiral galaxies, starbursts, or ultra-luminous infrared galaxies (ULIRGs)).
The difference in colors can be again interpreted as due to 4.5~$\mu$m excess at the S nucleus.

\subsection{MIR Spectrophotometry Modeling\label{MIR_SED_modeling}}

It is very likely that the S nucleus is composed of multiple components with different characteristics.
This is because the spectrum of the S nucleus at 0\farcs 36 scale is remarkably different from the spectra at arcsec resolution \citep{ds10}.
Within the 0\farcs 36 aperture, the S nucleus shows much deeper silicate absorption without detectable PAH.
The PAH features are evident in the IRS at $\sim 3$\arcsec~scale (\citealt{ds10,ps10}, this work), the wider-slit T-ReCS (1\farcs 3-wide; \citealt{lira08}), and the TIMMI2 (1\farcs 2-wide; \citealt{siebenmorgen04,mh06}) spectra.
This difference is probably because circumnuclear star-forming regions dominate the MIR flux within the larger apertures.
In terms of flux level, all the larger-aperture spectrophotometries are comparable to our IRAC 8.0~$\mu$m nuclear photometry, but the T-ReCS 8.0~$\mu$m flux within the 0\farcs 36 aperture is as small as 30\% of the IRAC flux.

We constructed simple composite SED models for the S nucleus with and without AGN to examine how additional AGN contribution can better reproduce the MIR observations compared to a pure (but absorbed) starburst model.
We fitted the models to both the high spatial-resolution T-ReCS spectrum, in particular its deep silicate absorption, and the nuclear IRAC and IRS spectrophotometries, in particular its prominent PAH features.
For simplicity, only two components are in the composites.
One of them needs to be compact and heavily absorbed to reproduce the very deep silicate absorption of the T-ReCS spectrum.
Another one represents the extended circumnuclear star-forming region which appears to dominate the PAH feature.
We made two such composite models.
The AGN-starburst composite model assumes a heavily absorbed AGN in addition to surrounding star-forming regions at $\sim 3$\arcsec~scale.
The starburst-starburst composite model assumes heavily and mildly absorbed starburst regions.
Emission from the heated dusts associated with the compact and heavily absorbed starburst component is not considered in the fitting.
As a reference, we also constructed a single-component absorbed starburst model without an AGN.
Although \cite{ah12} already did similar but more sophisticated SED modeling over the whole aperture IRS spectrum, ours is improved in the following three points.
Firstly, we include the IRAC 4.5~$\mu$m flux, below the IRS wavelength coverage, in the fitting so that the bluer SED at $<6$~$\mu$m of the S nucleus can be clearly traced.
The IRAC 3.6~$\mu$m flux is not included in the fitting because the flux is usually contaminated by stellar emission and relative contribution of the stellar component with respect to non-stellar component varies from one star-forming region to another.
Secondly, we utilize the nuclear spectrophotometry.
\cite{ah12} used a very large aperture IRS spectrum that includes both nuclei and their surroundings.
Given the only mild differences between the N and S nuclei at the IRS spatial resolution (\S \ref{infrared_agn_signature_results}), it is essential to separately analyze the nuclear spectra.
Thirdly, we require our model to fit the high-spatial-resolution T-ReCS spectrum while simultaneously fitting the IRAC and IRS spectrophotometry.

We fit the T-ReCS spectrum at 8.0--12.9~$\mu$m (\S \ref{trecs_data_analysis}) and the IRAC 4.5~$\mu$m--IRS spectrophotometries at 4.5--14.5~$\mu$m (\S \ref{irac_iras_fluxmatch}).
For reference, we also fit only the IRS spectra (5.2--14.5~$\mu$m) to evaluate the effect of including the IRAC 4.5~$\mu$m photometry.
We employed a standard $\chi^2$ minimization technique implemented in the MPFIT software package \citep{mpfit}.
For the starburst component, we adopt the LIRG SED template of NGC~6090 from \cite{swire} since the IRS starburst template of \cite{brandl06} does not completely cover the IRAC 4.5~$\mu$m channel.
For the AGN component, we adopt a pure power-law SED ($f_{\nu} \propto \nu^{\alpha}$) with a power-law index ($\alpha$) of $-0.5$.
The colors/flux ratios of this SED are well within the expected ones for AGNs (Figure~\ref{fig_irac_color_color}), and, as we show later, this intrinsically bluer power-law fits the observed IRAC SED when heavily extinct to fit the deep silicate feature of the T-ReCS spectrum.
For the extinction curve, we adopt three different types and compare the results because, as we demonstrate later, a slight difference in silicate absorption profile can result in quite different best-fit parameters with similarly good statistics for the same T-ReCS data.
This strongly affects our fit to the IRAC and IRS data covering wider wavelengths.
We cannot identify more appropriate extinction curve without having good anchor point(s) of the fit away from the very deep silicate absorption.
Due to the same problem, we fit the two sets of spectrophotometry data one by one because fit to the T-ReCS spectrum at 8.0--12.9~$\mu$m hardly constrains our fit around 4.5~$\mu$m on the IRAC and IRS spectrophotometries.
Therefore, we first fit the T-ReCS spectrum with each of the three extinction curves by two SED models (either starburst or AGN).
During the fit, we fix the redshift, and fit the SED scaling factor and extinction (represented by optical depth at 9.7$~\mu$m, $\tau_{\rm 9.7\mu m}$).
We then subtract the fitted model from the IRAC and IRS data, and fit the residual spectra with the starburst SED by using the same extinction curve.
We again fix the redshift, and fit the SED scaling factor and $\tau_{\rm 9.7\mu m}$.
We assume a screen geometry for both components.
We do not consider an effect of radiation from heated dust, although such a radiation is expected for a compact dusty region surrounding the central energy sources (either starburst or AGN).

We employ the following three extinction curves in the fitting.
The first one is a theoretical extinction curve of \cite{ct06} (hereafter, CT06) for the Galactic center.
The second is from a spectral fitting code PAHFIT \citep{pahfit}.
\cite{pahfit} introduced a hybrid curve consisting of an observed 9.7~$\mu$m silicate absorption profile, an empirical 18~$\mu$m silicate absorption profile, and a power-law component.
The 9.7~$\mu$m silicate feature is taken from \cite{kemper04} based on the observation toward the Galactic center.
This curve is found to work very well on IRS spectra of normal and star-forming galaxies (\citealt{pahfit}; see also, e.g., \citealt{ds10b,ps10,ds11}).
The third one uses the same function of PAHFIT but with a different parameter set to reproduce the updated extinction curve toward the Galactic center by \cite{fritz11}.
The curve of \cite{fritz11} shows stronger silicate absorption than that of CT06, although both curves are for the Galactic center.
For our fitting purpose, we modified parameters of the PAHFIT extinction curve to reproduce the very high-resolution extinction curve of \cite{fritz11}.
We set the power-law slope of the continuum extinction curve zero, and relative strength of the power-law component with respect to the 9.7~$\mu$m silicate feature strength ($\beta$; \citealt{pahfit}) 0.155.
Because the extinction curve of \cite{fritz11} is almost proportional to that of CT06 below the silicate feature, we replaced short side of the extinction curve ($<7.3$~$\mu$m) with the scaled curve of CT06.
This gives smaller $A_{\rm V}/\tau_{9.7}$ (5.8) than that of CT06 ($A_{\rm V}/\tau_{9.7}=9$).
The three extinction curves are compared in Figure~\ref{fig_extinction_curve}.
Results of the fits are summarized in Figure~\ref{fig_sedmodeling} and Table~\ref{tab_sed_fitting_results}.
As we show below, the original PAHFIT extinction curve gives much poorer fits, therefore the best-fit SED models are not shown with this curve in Figure~\ref{fig_sedmodeling}.

\subsubsection{Absorbed Single-Starburst Model}

In the absorbed single-starburst model (Figure~\ref{fig_sedmodeling} top), we modeled the IRAC 4.5~$\mu$m--IRS spectrophotometry of the S nucleus only with the absorbed LIRG SED template.
Although the model reproduces the IRS spectrum reasonably well, neither the 4.5~$\mu$m excess nor deep silicate absorption at the 0\farcs 36 scale can be reproduced.
Reduced $\chi^2$ are about 4--6 for all three extinction curves, and the IRAC 4.5~$\mu$m flux contributes most to the $\chi^2$.

\subsubsection{AGN-Starburst Composite Model\label{agn_starburst_composite_model}}

In the AGN-starburst composite model (Figure~\ref{fig_sedmodeling} middle), we found a good fit, although not statistically satisfactory, to the T-ReCS spectrum with the AGN template by using the extinction curve of CT06 (reduced $\chi^2=1.4$) and our modified PAHFIT extinction curve (reduced $\chi^2=1.2$).
On the other hand, the original PAHFIT extinction curve gives poor fit (reduced $\chi^2=3.6$) because the curve always produces asymmetric extinction at both ends of the silicate absorption profile, while the observed profile is more symmetric than the curve.
We note that the best-fit parameters are significantly different for different extinction curves ($\tau_{\rm 9.7}$=9.4 and 12.7 for the extinction curve of CT06 and our modified PAHFIT extinction curve, respectively).
As explained earlier, this is due to slightly different profile of the silicate absorption feature and absence of fit anchor point(s) outside the feature.
For the fit to the IRAC 4.5~$\mu$m--IRS spectrophotometers, both extinction curve of CT06 (reduced $\chi^2=1.4$) and our modified PAHFIT extinction curve (reduced $\chi^2=0.8$) resulted in $\tau_{\rm 9.7}\simeq 0.5$--0.7.
We note that the IRAC 4.5~$\mu$m flux dominates the $\chi^2$ in a fit with the extinction curve of CT06 in a sense that the fit under-predicts the observation.

\subsubsection{Starburst-Starburst Composite Model\label{starburst_starburst_composite_model}}

In the starburst-starburst composite model (Figure~\ref{fig_sedmodeling} bottom), we found that a heavily absorbed LIRG SED also fits the T-ReCS spectrum (reduced $\chi^2=1.2$ for the extinction curve of CT06 and our modified PAHFIT extinction curve).
This is because the silicate absorption profile of the extinction curve essentially dominates the shape of the absorbed SED within the T-ReCS wavelength coverage.
Again, the original PAHFIT extinction curve gives poor fit (reduced $\chi^2=2.7$) because of the same reason for the case of the AGN-starburst composite model.
The fits to the IRAC 4.5~$\mu$m--IRS spectrophotometries (reduced $\chi^2 \simeq 4$--6) are much worse than those of the AGN-starburst composite model (reduced $\chi^2 \simeq 0.8$--1.4), because contribution to the 4.5~$\mu$m flux from the heavily absorbed LIRG template is negligibly small.

\subsubsection{Summary and Implication of the SED Modeling Results\label{summary_MIR_SED_modeling}}

We found that the AGN-starburst composite model fits the MIR observations much better than the starburst-starburst composite model since AGN contribution can reproduce the 4.5~$\mu$m excess.
In particular, the modified PAHFIT extinction curve gives better fit to the 4.5~$\mu$m flux, and we prefer the results with this extinction curve, i.e., Figure~\ref{fig_sedmodeling} right-middle.
This model can also roughly explain the NICMOS 2.2~$\mu$m flux of the unresolved nuclear component, although the flux is not included in the fit (Figure~\ref{fig_sedmodeling_best}).
In our best-fit model with the modified PAHFIT extinction curve, the AGN contributes $\simeq 24$\% and $\simeq 2$\% of the 6 and 24~$\mu$m fluxes, respectively.
As for the optical depth on the AGN component, we use both results, $\tau_{\rm 9.7}=9.4\pm 0.4$ and $12.7\pm 0.5$ from the extinction curve of CT06 and our modified PAHFIT extinction curve, respectively, and adopt the possible range of the optical depth of $\tau_{\rm 9.7}=9$--13, since the two extinction curves provide statistically equally good fit to the T-ReCS spectrum.

We estimate the column density toward the AGN to be on the order of $N_{\rm H} \sim10^{23}$ cm$^{-2}$.
We assumed a screen geometry of the dusty region.
The optical depth at 9.7~$\mu$m over the AGN continuum, $\tau_{\rm 9.7}=9$--13, corresponds to $A_{\rm V}\sim 80$ mag by using the conversion factors of $A_{\rm V}$/$\tau_{\rm 9.7} = 9$ and $5.8$ for extinction curves of CT06 and our modified PAHFIT, respectively.
Further assuming a standard conversion factor for the solar neighborhood, $N_{\rm H}$/$A_{\rm V} = 1.9 \times 10^{21}$ cm$^{-2}$ mag$^{-1}$ \citep{bohlin78}, the neutral Hydrogen column density toward the AGN is estimated to be $N_{\rm H} \simeq 1.5 \times 10^{23}$ cm$^{-2}$.

We can also estimate the star formation rate (SFR) and IR luminosity of the S nucleus.
By subtracting the synthetic 8.0$~\mu$m flux of the AGN component (16~mJy), we estimate 8~$\mu$m flux from the starburst component of our best AGN-starburst composite model is 35~mJy.
We then estimate SFR to be 0.43 M$_{\odot}$ yr$^{-1}$ by using calibration of \cite{zhu08}, which is derived for young (10--100~Myr continuous burst) starburst with Salpeter initial mass function (0.1--100 M$_{\odot}$) based on IR (8--1000~$\mu$m) luminosity--SFR conversion and monochromatic 8~$\mu$m luminosity--IR luminosity correlation.
This SFR corresponds to $L_{\rm IR} \simeq 2.5\times10^{9}$ L$_{\odot}$.
\cite{zhu08} noted that correlation between MIR and IR luminosities are essentially the same for AGN-starburst composite galaxies, AGNs, and star-forming galaxies.
Therefore, the IR luminosities of the AGN and the S nucleus, including the AGN and the circumnuclear starburst component, are estimated to be $\simeq 1.1\times10^{9}$ L$_{\odot}$ and $3.6\times10^{9}$ L$_{\odot}$, respectively\footnote{\cite{sakamoto14} referred to this paper in its submitted form for the FIR luminosities. The quoted luminosities are $\simeq 2.9\times10^{10}$ L$_{\odot}$, $\simeq 1.5\times10^{10}$ L$_{\odot}$, and $\simeq 5.0\times10^{9}$ L$_{\odot}$ for the N nucleus, the S nucleus, and the AGN in the S nucleus, respectively. During the revision of this paper, we revised the numbers as shown here.}.
We note that the AGN luminosity is highly uncertain because the correlation between the MIR and IR luminosities is established for field AGNs found in a blank-sky survey \citep{zhu08}, and it is not clear if the same correlation applies to the heavily absorbed AGN at the S nucleus.
For comparison, the IR luminosity of the N nucleus estimated in the same way from the observed 8.0$~\mu$m flux is $\simeq 1.5\times10^{10}$ L$_{\odot}$.
The total infrared luminosity at 3--1100~$\mu$m (TIR) of the N nucleus is also estimated from the whole-aperture TIR luminosity by \cite{engelbracht08}, who used both IRAC and MIPS \citep{mips} 24~$\mu$m, 70~$\mu$m, and 160~$\mu$m fluxes.
Since the IRAC colors of the N nucleus are very similar to that of the rest of the galaxy, and the S nucleus that shows distinct colors consists of only a few percent of the whole aperture flux (\S \ref{irac_data_analysis}), we scale the whole-aperture TIR luminosity according to the nuclear fluxes at 8.0~$\mu$m.
We then found $L_{\rm TIR}=$(2.0--2.4) $\times 10^{10}$ L$_{\odot}$ for the N nucleus.

\section{X-RAY DATA}\label{x_ray_analysis}

We analyzed archival data of two {\it Chandra} observations of NGC~3256 to search for X-ray indications of the presence of an AGN
in the S nucleus.
The superb spatial resolution of {\it Chandra} enables us to
extract X-ray spectrum of the S nucleus separated from other emission components in the host galaxy.
The two {\it Chandra} observations are summarized in Table~\ref{tab_chandra_log}.
The observations performed on 2000 Jan 5 and 2003 May 23 are referred to as the first and second observations,
respectively, hereafter. NGC~3256 was located on the back illuminated 
CCD chip ACIS-S3 in both observations.
The {\it Chandra} Interactive Analysis of Observations (CIAO) software package version 4.6 combined with 
the latest calibration database (CALDB) version 4.6.3 was used to analyze the {\it Chandra} data.
The data were reprocessed to generate level=2 event files using the latest calibration.

We made a light curve for a source free region in the same CCD chip to examine the stability of background and discarded
time intervals showing high background.
The background was stable in the second observation, while the background rates were
relatively high in the first. The resulting exposure times after discarding high background intervals are 16.2 and 27.2~ksec
for the first and second observations, respectively, totaling $\simeq 2.7$ times more exposure time than that in \cite{lira02}.
X-ray spectra of the S nucleus were extracted using a circular region with a radius of 2.2 pixels or 1\farcs 1.
Background spectra were made using a circular region with a radius of
5\farcs 2 (first observation) or 5\farcs 9 (second observation)
that does not contain noticeable point sources near NGC 3256
and were subtracted from the source spectra.
The net counts after the background subtraction are shown in Table~\ref{tab_chandra_log}
and the background subtracted spectra are presented in Figure~\ref{chandra_spec}.
The spectra of the S nucleus were binned so that each bin contains at least one count.

Spectral fits were performed by using XSPEC
version 12.8.2. We applied a maximum likelihood method to fit the spectra using 
the modified version of $C$ statistic \citep{cash79}, in which a Poisson distribution is assumed for numbers of counts in each bin.
While the absolute value of $C$ does not provide goodness of fits, $\Delta C$ can be used to examine relative goodness or to generate confidence intervals.
The errors are at the 90\% confidence range for one parameter of interest ($\Delta C = 2.7$).
The Galactic absorption $N_{\rm H} = 9.35\times 10^{20}$ cm$^{-2}$ \citep{kalberla05}
calculated by using the tool {\tt nh} in FTOOLS was applied to all the models examined below.
The {\tt phabs} model in XSPEC was used
for photoelectric absorption, in which the 
cross sections from \cite{balucinska92} with a He cross section in \cite{yan98} are used.
All the spectral components except for the Galactic absorption were assumed to be emitted/absorbed at the source redshift.

\section{X-RAY RESULTS}\label{x_ray_results}

We fitted the spectra from the two observations simultaneously. Since the photon statistics are limited, we assumed same spectral parameters for
the two spectra.
A constant factor was multiplied to the model to represent variability between the two observations. 
This factor was fixed at unity for the first observation and left free for the second observation.

\subsection{Power-Law Models}\label{powerlaw_model}

A simple power law modified by absorption intrinsic to the source (model A) was examined first. The resulting parameters are summarized in Table~\ref{tab_chandra_pl_fit_results}, and 
the photon index and absorption column density of
$-0.42^{+0.67}_{-0.27}$ and $<9.7\times10^{21}$ cm$^{-2}$ (90\% confidence upper limit), respectively, were obtained. Although this model represents the overall shape of the spectra, the best-fit photon index is extremely flat.
If the photon index is fixed at a steeper value 1.8, which is typically observed in Seyfert galaxies (e.g., \citealt{dadina08}),
the $C$ statistic becomes worse by $\Delta C$ = 17.7 (Model B).

An apparently flat spectrum is often explained by a reflection dominated spectrum or a combination of heavily absorbed and lightly absorbed continua, and these possibilities are tested below.
The former case is expected if the primary X-ray emission is obscured by a large amount of matter exceeding a column density of $\sim10^{24}$ cm$^{-2}$ and if emission scattered by medium surrounding the X-ray source dominates observed spectra. In oder to represent a spectrum dominated by reflection by cold matter, the {\tt pexrav} model \citep{magdziarz95} in XSPEC modified by intrinsic absorption was used. The incident spectrum was assumed to be power law with an exponential high energy cutoff. The photon index and the cutoff energy were fixed at 1.8 and 300 keV, respectively, since these parameters were not well constrained from the observed spectra. A reflector of a slab shape 
with an inclination angle of 60$^\circ$ was assumed, where 0$^\circ$ corresponds to face on. 
The abundance of the reflector was assumed to be solar, where the abundance table of \cite{anders89} was used.
The reflection normalization factor (``{\tt ref\_refl}'' parameter) was fixed at $-1$ so that only the reflected continuum emission is obtained. 
This model (Model C) represents the observed spectral shape, and the best-fit parameters are shown in Table~\ref{tab_chandra_pl_fit_results}, though the absence of Fe-K emission line is not compatible with this refection dominated model as discussed in \S \ref{x_ray_evidence}.

Another possibility to explain the very flat continuum is a multicomponent model consisting of heavily and lightly absorbed continua. We assumed a power law continuum with a fixed photon index of 1.8 as an incident spectrum and the absorption column densities of
$N_{\rm H, 1}$ and $N_{\rm H, 2}$ for light and heavy absorbers, respectively.
This model is expressed as
 $e^{-\sigma(E) N_{\rm H, 1}} ~[f e^{-\sigma(E) N_{\rm H, 2}} + (1-f)]~A~E^{-\Gamma}$,
where $E$, $\sigma(E)$, $f$, $A$, and $\Gamma$ are photon energy, 
photoelectric absorption cross section, fraction of continuum emission absorbed by $N_{\rm H, 2}$, normalization of power law, and
photon index, respectively (Model D). The observed spectra are well represented by this model.
The result of this fit is also in Table~\ref{tab_chandra_pl_fit_results}.
The best-fit model is shown in Figures \ref{chandra_spec} and \ref{chandra_ufsp}.
The measured column density of the heavy absorber ($7^{+19}_{-3} \times10^{22}$ cm$^{-2}$) is consistent with the MIR measurement of an AGN absorption from the AGN-starburst composite model ($\simeq 1.5 \times 10^{23}$ cm$^{-2}$) (\S \ref{summary_MIR_SED_modeling}).

The constant factors multiplied to the models obtained from the fits are summarized in Table~\ref{tab_chandra_pl_fit_results}. The best-fit values for the models examined above are in the range from 0.92 to 0.95, and the error ranges contain unity. Therefore, flux variability between the two observations is not significant.
Observed fluxes and luminosities corrected for absorption in the 2--10 keV band are also shown in Table~\ref{tab_chandra_pl_fit_results} except for the model B, for which
$C$ statistic was significantly worse than those for other models. 

The observed spectra do not show an indication of an Fe-K emission line. We calculated an upper limit on the EW of a fluorescent 
Fe-K line at 6.4 keV by adding a Gaussian spectral component to the models examined above.
The center energy was fixed at 6.4 keV. The line width $\sigma$ is assumed
to be 10 eV, since previous observation of type 2 AGNs show that Fe-K line
width is much narrower than the energy resolution of ACIS \citep{shu11}.
The upper limit on the line EW is shown in Table~\ref{tab_chandra_pl_fit_results}.

\subsection{Thermal Emission Models}\label{thermal_model}

We also examined a thermal emission model. The APEC model \citep{smith01} was used as emission from optically-thin plasma in collisional ionization equilibrium.
A combination of multiple components, one of which is heavily absorbed, is required to explain the apparent flat spectrum. 
We therefore examined a 
model consisting of unabsorbed and absorbed APEC components
represented by the expression
{\tt apec}$(T_1) + e^{-\sigma(E) N_{\rm H}}$ {\tt apec}$(T_2)$, where $T_1$ and $T_2$ 
are temperatures for unabsorbed and absorbed APEC components, respectively (Model E). 
Only lower bounds for allowed temperature ranges for these two components were obtained as 1.2 keV and 9.4 keV, respectively. The intrinsic column density for the absorbed component was obtained to be 
$N_{\rm H} = 5.1^{+3.3}_{-1.9}\times10^{22}$ cm$^{-2}$. This model resulted in the value of $C$ statistic 77.6 for 88 degrees of freedom and describes the shape of the observed spectra, where the plasma temperatures $kT_1$
and $kT_2$ were pegged at 64 keV.
X-ray spectra of starburst galaxies generally show emission from thermal
plasma with a temperature $kT=1$ keV or less
\citep{ptak99,strickland04,tullmann06}, and the
high temperatures obtained from the fits are unusual. We therefore
examined whether the presence of low temperature ($kT<1$ keV) plasma is
consistent with the observed spectrum by multiplying extra absorption to
the model examined above. This trial model is
expressed by
$e^{-\sigma(E) N_{\rm H, 1}} [$
{\tt apec}$(T_1) + e^{-\sigma(E) N_{\rm H, 2}}$ 
{\tt apec}$(T_2) ]$, 
and we found that $kT_1$ of 1 keV or less is allowed
for the component representing the low energy part of the spectra (Model F).
If $kT_1$= 1 keV is assumed, the best-fit $N_{\rm H, 1}$ is $1.3\times10^{22}$ cm$^{-2}$, and $C$=77.3 is obtained for 88 degrees of freedom. 
The resulting spectral parameters, observed fluxes, and absorption corrected luminosities
for the thermal models are summarized in Table~\ref{tab_chandra_thermal_fit_results}.

\section{DISCUSSION}

\subsection{Evidence of AGN in the S Nucleus\label{agn_evidence}}

\subsubsection{In Infrared\label{ir_evidence}}

\begin{figure}
\plotone{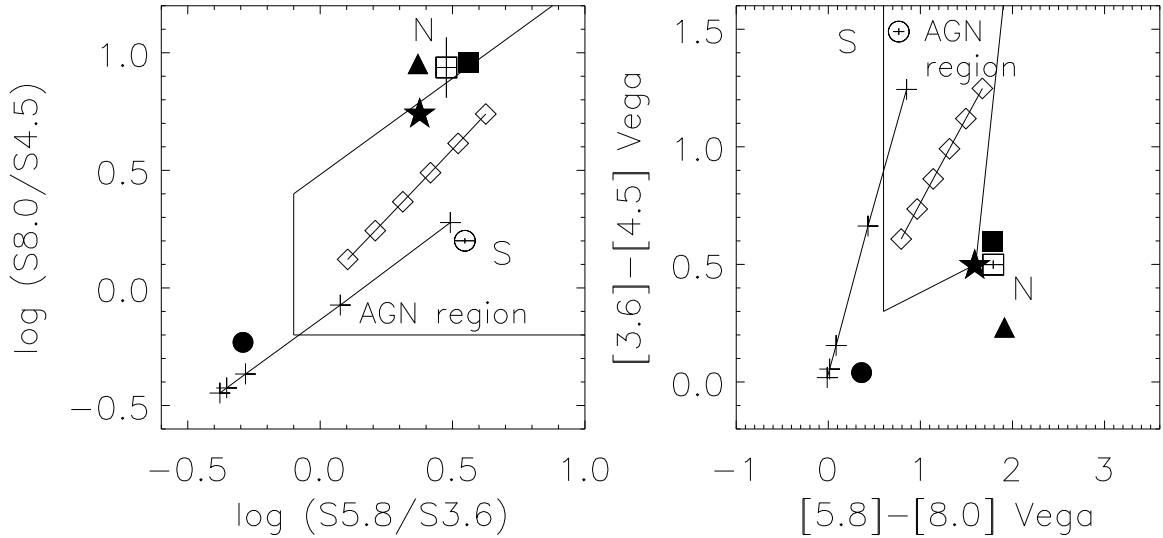}
\caption{
The IRAC color-color diagrams of the N (open square) and S (open circle) nuclei with boundaries for AGN selection.
Error bars are $\pm 1\sigma$.
Also shown in filled symbols are rest frame colors/flux ratios of ULIRG IRAS~22491-1808 from \cite{swire} (square), starburst M~82 from \cite{swire} (star), a template of normal star-forming spiral galaxy from \cite{dale02} (triangle), and a representative elliptical galaxy from \cite{swire} (circle).
Connected diamonds are for pure power-law SEDs, and each triangle corresponds to a power-law index of $\alpha$ from $-0.5$ (lower left) to $-3.0$ (upper right) in steps of 0.5.
Connected pluses represent colors/flux ratios for pure blackbody SEDs, and each plus corresponds to a blackbody temperature of 600~K (upper right), 1000~K, 3000~K, 6000~K, and 10000~K (lower left).
The boundaries for AGN selection are from \cite{donley12}.
\label{fig_irac_color_color}}
\end{figure}

\begin{figure}
\plotone{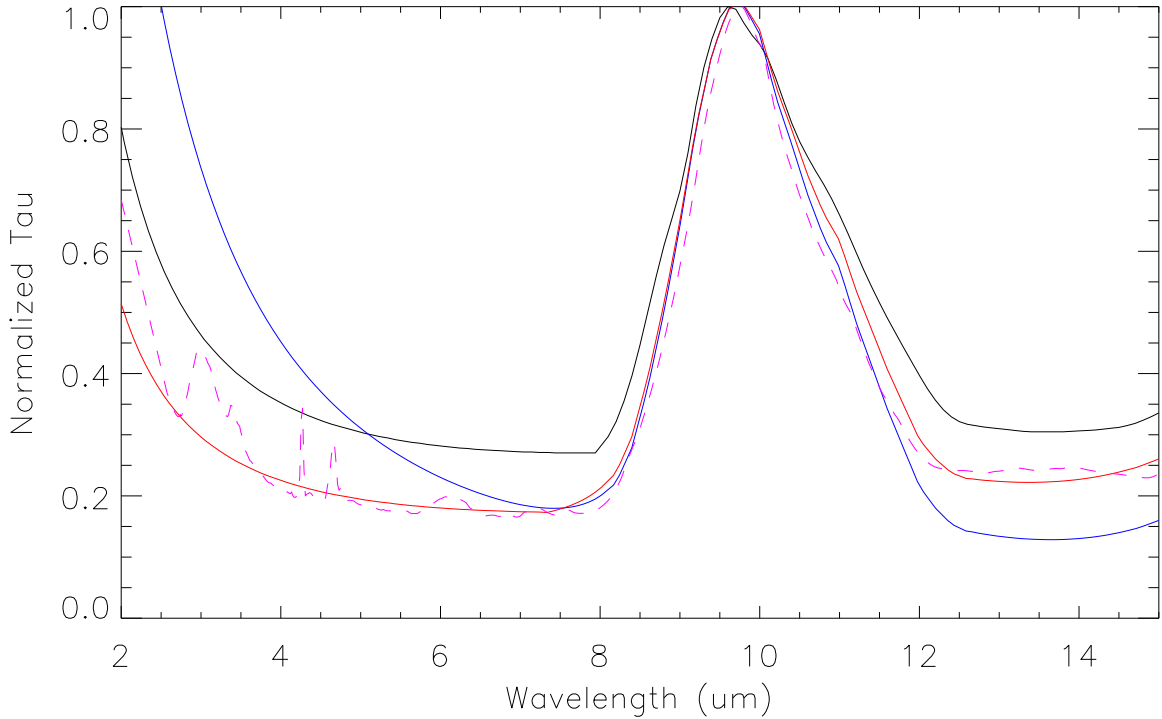}
\caption{Comparison of extinction curves used in our SED modeling.
We show the extinction curve of \cite{ct06} for the Galactic center (black solid line), the original PAHFIT \citep{pahfit} extinction curve (blue solid line), the original extinction curve of \cite{fritz11} (broken magenta line), and the modified PAHFIT extinction curve (red solid line) to reproduce the curve of \cite{fritz11}.
See the main text for the details of the modified PAHFIT extinction curve.
\label{fig_extinction_curve}}
\end{figure}

\begin{figure}
\plotone{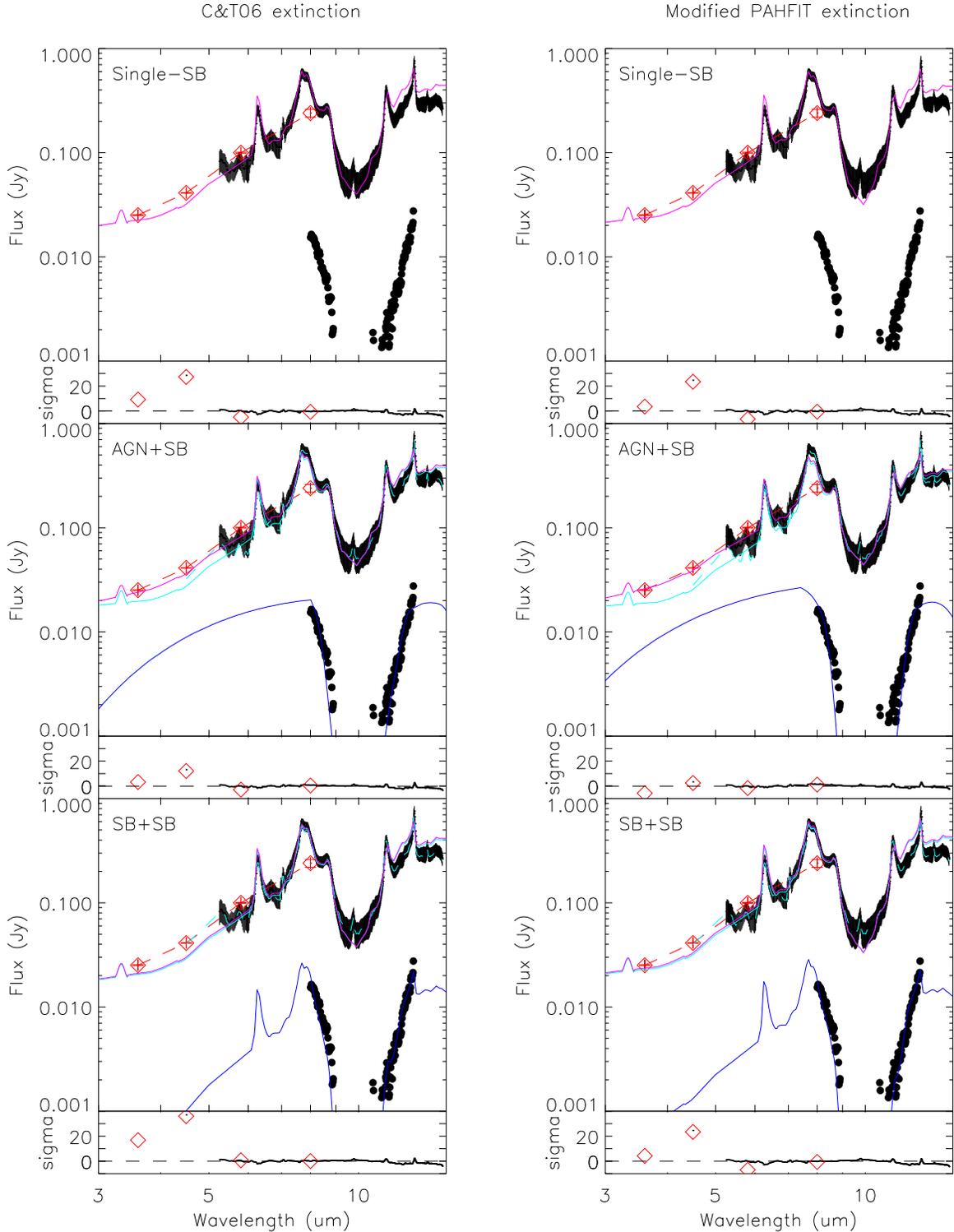}
\caption{
SED modeling results for the S nucleus.
The absorbed single-starburst model (Single-SB), the AGN-starburst composite model (AGN+SB), and the starburst-starburst composite model (SB+SB) are shown in the top, middle, and bottom rows, respectively.
Results with the extinction curve of \cite{ct06} and the modified PAHFIT extinction curve are shown in the left and right columns, respectively.
The adjusted IRAC photometries as shown in Figure~\ref{fig_irac_irs_sed} (connected red diamonds with $\pm 1\sigma$ error bars), the IRS (with $\pm 1\sigma$ error bars; black) and T-ReCS spectrophotometries (black) are shown in all panels.
For the absorbed single-starburst model, we overlay the best-fit models with magenta solid lines.
For both the AGN-starburst and starburst-starburst composite models, we overlay the best fit models for the T-ReCS (blue), residual 4.5~$\mu$m--IRS spectrophotometries (observed$-$best-fit T-ReCS model) (cyan), and sum of the two components (magenta).
The sigma ((observation$-$model)$/$observation noise) spectra of the IRS spectrophotometries (black) and synthetic IRAC fluxes from the best-fit models (red diamonds) are shown for each panel.
\label{fig_sedmodeling}}
\end{figure}

\begin{figure}
\plotone{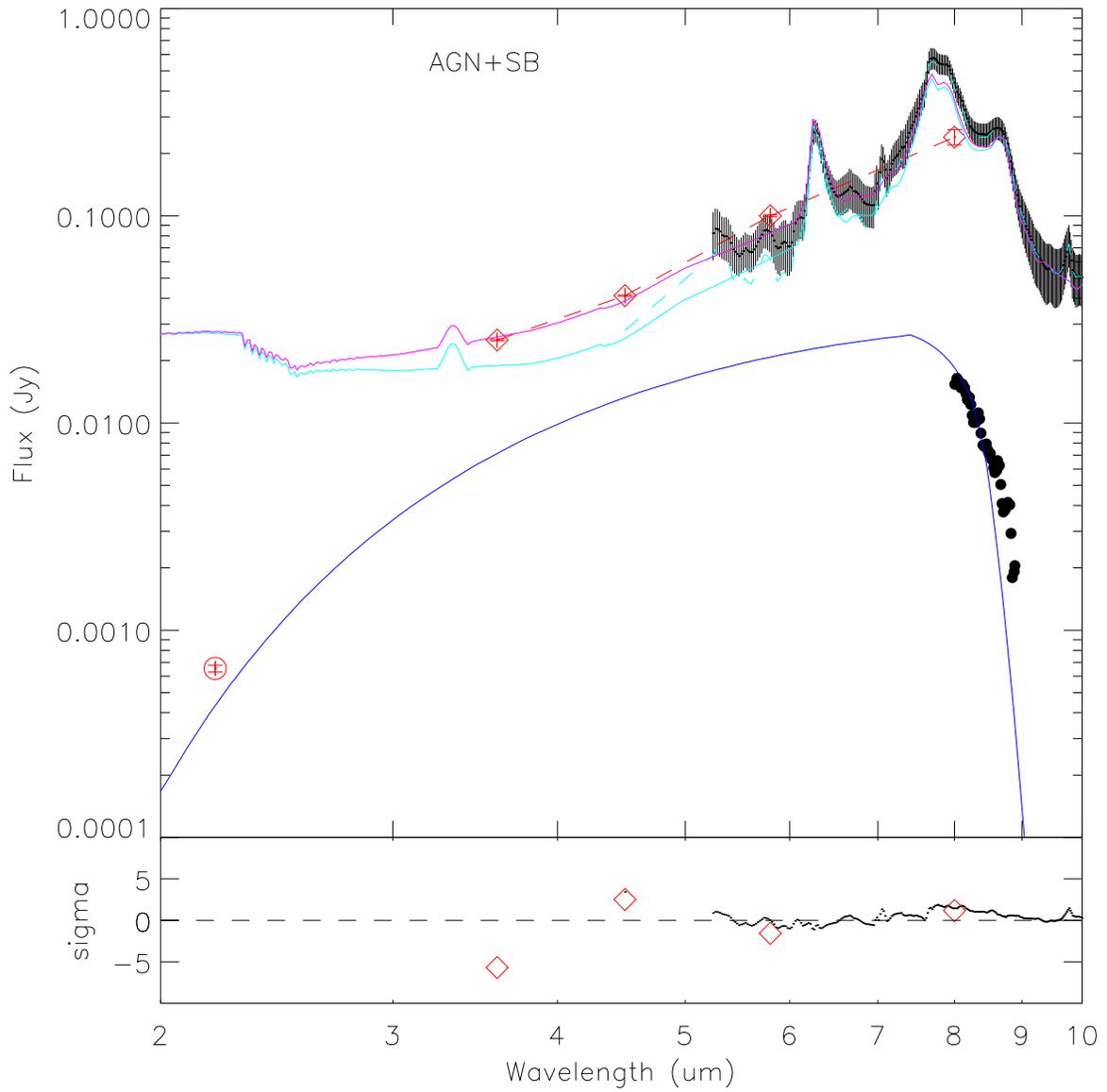}
\caption{
The best SED modeling result for the S nucleus with the AGN-starburst composite model and the modified PAHFIT extinction curve at 2.0--10~$\mu$m.
The same colors and symbols of Figure~\ref{fig_sedmodeling} are used.
The NICMOS photometry at 2.2~$\mu$m of the unresolved core is also shown with a $\pm 1\sigma$ error bar (red circle).
\label{fig_sedmodeling_best}}
\end{figure}

\begin{center}
\begin{figure}
\includegraphics[angle=270,scale=0.85]{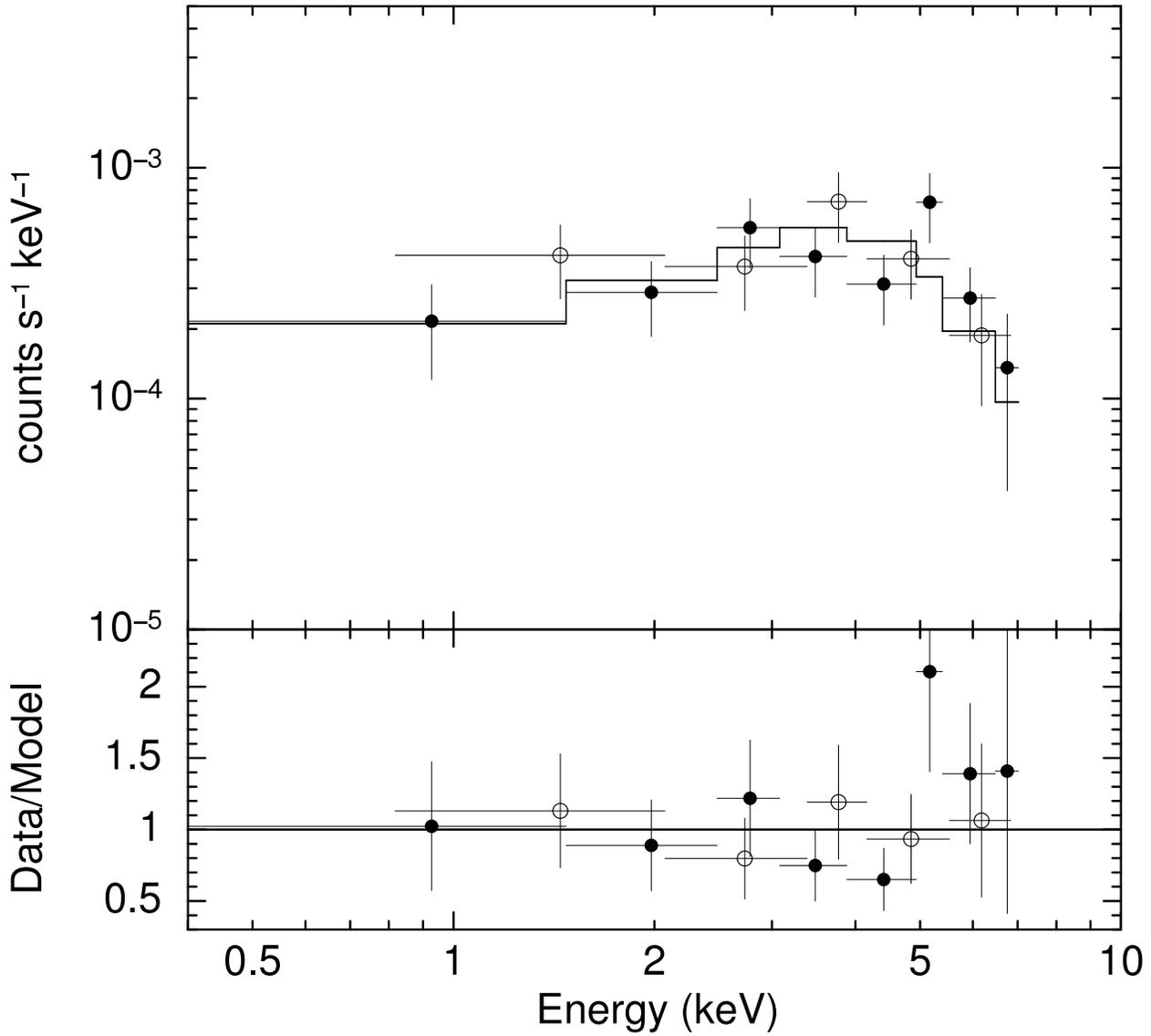}
\caption{Observed {\it Chandra} spectra of the S nucleus for the first (open circles) and second (filled circles) observations. Data are binned for presentation purpose. (Upper panel) Observed spectra and the best-fit model (Model D, solid line). Only the model for the second observation is shown for clarity. (Lower panel) Data/Model ratio.\label{chandra_spec}}
The error bars are for $\pm 1\sigma$.
\end{figure}
\end{center}

\begin{center}
\begin{figure}
\includegraphics[angle=270,scale=0.85]{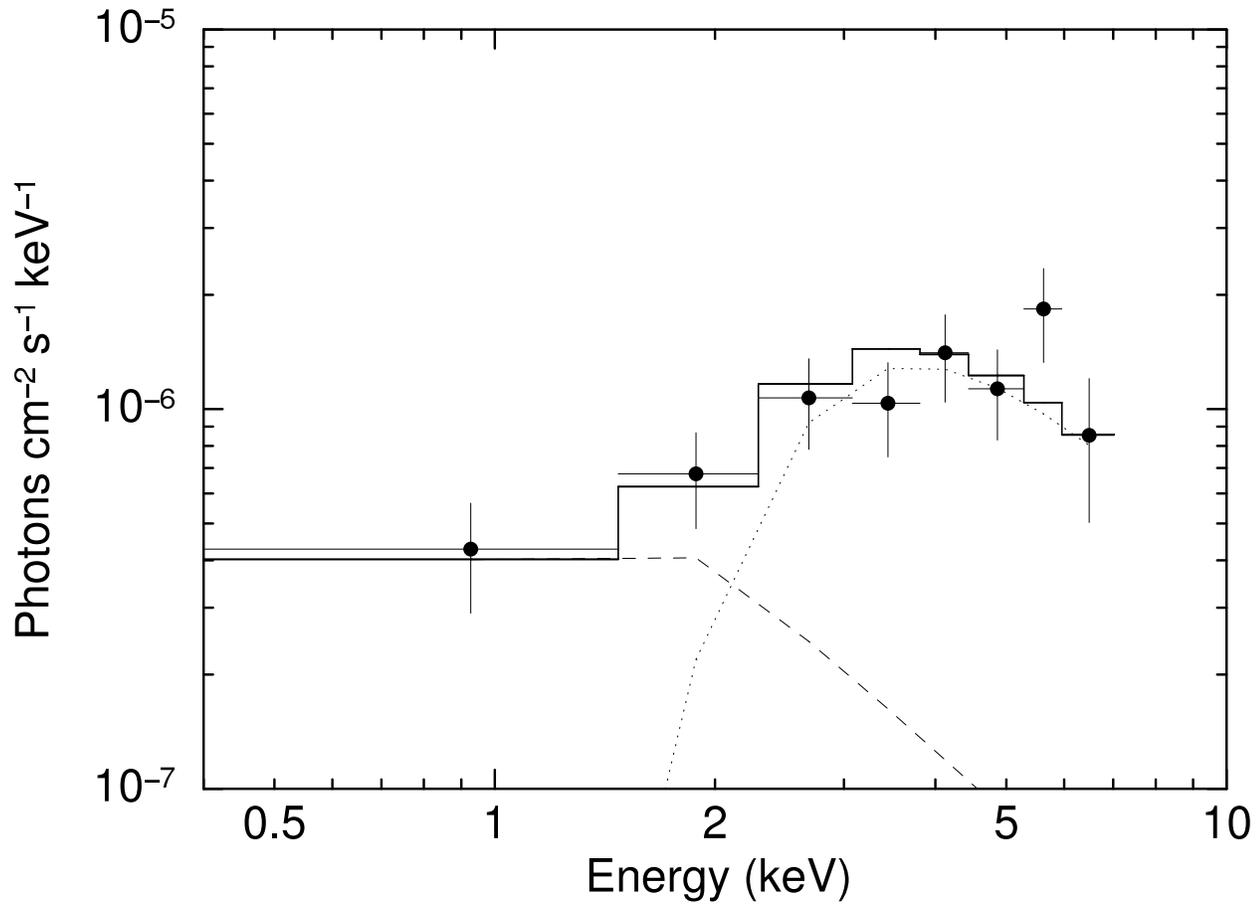}
\caption{Unfolded {\it Chandra} spectrum of the S nucleus. The responses of detector and telescope are unfolded.
The spectra from the two observations are combined for presentation purpose.
The error bars are for $\pm 1\sigma$.
The best-fit model (Model D, solid line),
lightly (dashed line), and heavily (dotted line) absorbed components are shown for comparison.
\label{chandra_ufsp}}
\end{figure}
\end{center}

The IRAC colors at the S nucleus indicate a non-starburst-like SED (\S \ref{infrared_agn_signature_results}).
Specifically, the excess of the 4.5~$\mu$m flux makes 8.0~$\mu$m/4.5~$\mu$m flux ratio smaller and [3.6]$-$[4.5] color larger in both diagnostic diagrams in Figure~\ref{fig_irac_color_color}, bringing the S nucleus away from starburst-dominated galaxies in the observed colors.
Such colors are found only at the S nucleus in the IRAC images (\S \ref{infrared_morphology}).
The MIR spectrum of the flux ratio between the two nuclei (Figure~\ref{fig_irs_irac_ratio}a) can be used to examine presence of an AGN in a way least dependent on the SED templates.
It shows a complicated trend of both the excessed 4--6~$\mu$m flux and the deeper 9.7~$\mu$m absorption at the S nucleus.
The ratio changes in a way different from the dust extinction curves (Figure~\ref{fig_extinction_curve}) and, therefore, the two nuclei must have different intrinsic SED shapes.
This conclusion is independent of the detailed shape of the extinction curve.
Such an excess at 4--6~$\mu$m cannot be reproduced only from starburst-dominated SEDs because the observed 8.0~$\mu$m/4.5~$\mu$m flux ratio is smaller than, and the observed [3.6]$-$[4.5] color is larger (redder) than, those of starburst-dominated galaxies \citep{ruiz10}.
We demonstrated with our AGN-starburst composite SED model that the AGN contribution can reproduce the observed 4.5~$\mu$m enhancement (\S \ref{summary_MIR_SED_modeling}).

We discovered a very compact 2.2~$\mu$m source at the S nucleus, and its unresolved core (at 0\farcs 26 FWHM) seems a NIR counterpart of the compact (at 0\farcs 30 FWHM resolution) MIR core \citep{ah06b}.
The 2.2~$\mu$m core coincides with the compact MIR and radio sources at the S nucleus (\S \ref{infrared_morphology}).
Both the 2.2~$\mu$m core and the MIR core show a comparable source size.
Although this 2.2~$\mu$m component is not included in our SED fitting together with the T-ReCS spectrophotometries, the heavily extinct power-law AGN SED that is fitted to the MIR data roughly predicts the 2.2~$\mu$m flux (\S \ref{summary_MIR_SED_modeling}).
Therefore, it is very likely that the same power-law component, or hot blackbody-dominated component (see below), dominates the NIR--MIR SED of the nucleus at $<50$~pc scale.

Although our SED model fitting prefers a model with AGN, the nuclear IRAC colors can be also represented by a blackbody of $\simeq 600$~K (Figure~\ref{fig_irac_color_color}).
Although we neglected dust emission in our SED modeling of the starburst-starburst composite model (\S \ref{MIR_SED_modeling}), such dust emission might dominate the SED around 4.5~$\mu$m to account for the 4.5~$\mu$m excess.
\cite{verley07} studied SEDs of embedded star-forming regions by considering hot dusty shell with radiative-transfer calculation with {\it DUSTY} code \citep{dusty}.
No PAH emission is considered within {\it DUSTY}.
The young star clusters are modeled with the {\it starburst99} \citep{starburst99} stellar-population model.
Among the parameter space they explored, which was designed to cover a range of typical extragalactic H~{\sc ii} regions, the S nucleus is closer in IRAC colors to cases of temperature of the inner surface of the dusty shell $T_{\rm in}=600$~K and total visual extinction $A_{\rm V}=46.4$ mag.
A heavy 9.7~$\mu$m silicate absorption is also reproduced when the extinction is very high ($A_{\rm V}\sim 100$ mag).
In such configurations of the dusty shell, large extinction suppresses stellar SED at 3.6~$\mu$m, and the color becomes similar to a pure blackbody of $T=600$~K.
Luminosity of the 600~K blackbody component is $\sim 3\times 10^8$ L$_{\rm \odot}$ if the blackbody dominates the observed nuclear IRAC SED.
If we assume a shell of dusts emitting such blackbody, radius of the shell is $< 0.1$ pc.

MIR SEDs with dominating hot blackbody are unusual for star-bursting galactic nuclei.
Some types of star-forming objects are known to show SEDs similar to the IRAC SED at the S nucleus, but they are likely not responsible for the S nucleus for the following reasons.
Hot blackbody-dominated SEDs are found in blue compact dwarf galaxies \citep{hunt05} and ultra-compact H~{\sc ii} regions (e.g., \citealt{churchwell90,churchwell02}).
In blue compact dwarf galaxies, vigorous star formation is clearly seen in optical wavelength, and the MIR SEDs resemble a blackbody because of lack of prominent PAH emission due to low metallicity.
Because the S nucleus is most likely a nucleus of a massive disk galaxy under merging process (e.g., \citealt{english03,sakamoto14}) and the metallicity of the nuclear region is about solar (e.g., \citealt{sb95,lipari00}), the same explanation for blue compact dwarf galaxies unlikely applies to the S nucleus.
In addition, typical X-ray luminosity of such galaxies is about a few $10^{39}$ erg s$^{-1}$ \citep{kaaret11,thuan14}, which is about an order of magnitude fainter than the observed luminosity of the S nucleus.
Ultra-compact H~{\sc ii} regions are manifestations of newly formed massive stars that are still embedded in their natal molecular cloud, and hence dusty cocoon \citep{churchwell90}.
It is compact ($<$1~pc) and emits mostly at infrared from hot dust heated by central O star.
Strong silicate 9.7~$\mu$m absorption is sometimes seen.
Although they are among the most luminous FIR sources in the Galaxy and their MIR spectral characteristics are similar to those of the S nucleus, even the most luminous ultra-compact H~{\sc ii} regions in a classical sample of \cite{wood89} are $\sim 4 \times10^3$ times less luminous than that of the S nucleus at $\sim 10$~$\mu$m.
Their X-ray luminosities (typically a few $10^{33}$ erg s$^{-1}$; \citealt{tsujimoto06}) are also much fainter than the observed luminosity of the S nucleus.
Nuclear starburst galaxies and (U)LIRGs hardly show such hot dust component \citep{marshall07,armus07,dacunha10}.
In a sample of representative local star-forming ULIRGs, \cite{dacunha10} requires no component hotter than 250~K in their UV-FIR SED fitting.
\cite{armus07} pointed out from their detailed analysis of the infrared (1--1000~$\mu$m) SEDs for local ULIRGs with and without AGNs that detection of hot dust at $T\gtrsim 300$~K in a nuclear spectrum provides indirect evidence for a buried AGN.

In summary, our infrared analysis strongly prefers presence of an AGN at the S nucleus.
In order to better constrain the nuclear activities, sophisticated SED decomposition as \cite{marshall07} did or physical SED modeling with dust radiation-transfer consideration would be necessary.
Additional high spatial-resolution photometry data to fill a wavelength gap between NICMOS and $N$ band as well as spatially-resolved photometry of the S nucleus at longer infrared wavelength would greatly help such modeling and analysis.

\subsubsection{In X-Ray}\label{x_ray_evidence}

{\it Chandra} imaging of NGC~3256 shows an X-ray source at the position of the S nucleus. Its spatial extent is consistent with the PSF of {\it Chandra} \citep{lira02}. We first discuss possible origins of the X-ray emission assuming that X-rays are coming from one point source.
The X-ray spectra are extremely hard; if a simple power law model modified by intrinsic absorption is applied, the best-fit photon index of $-0.43$ is obtained (Model A). 
Such a flat spectrum is unusual for primary X-ray emission and suggests a contribution of reprocessed and/or absorbed emission. If 
the primary source is hidden behind optically thick matter with a column density greater than $N_{\rm H} \sim10^{24}$ cm$^{-2}$ and the observed spectrum is dominated by emission scattered by cold matter, the resulting spectrum becomes very hard. Such a spectrum is referred to as reflection dominated. This situation is observed in 
Compton-thick AGNs, for which an absorption column density exceeds $1.5\times10^{24}$ cm$^{-2}$
(e.g., \citealt{comastri04}). In this case, a strong fluorescent Fe-K emission line with an EW greater than $\sim 700$ eV is seen
(e.g., \citealt{guainazzi05,fukazawa11}). The observed spectra, however, do not show an indication of Fe-K emission line, and the upper limit on the EW is 190 eV (for model C).
This limit is inconsistent with a reflection dominated spectrum and the presence of a Compton-thick nucleus is ruled out.

A dual absorber model (Model D) also provided a good description of the spectrum. Such spectra are seen in AGNs obscured by Compton-thin matter. The fraction, $f=0.94$, of the continuum absorbed by a large column density ($7\times10^{22}$ cm$^{-2}$) is in the range typically observed in Seyfert 2s \citep{turner97,noguchi10}. The less absorbed emission component is often interpreted as emission scattered by ionized medium in the opening part of the putative obscuring torus (e.g., \citealt{turner97}).
The upper limit on the EW of an Fe-K emission line (550 eV) is consistent with that expected for the absorption column density of $7\times10^{22}$ cm$^{-2}$ (EW$\simeq$ 50--150 eV; \citealt{turner97,guainazzi05,fukazawa11}). The luminosity in the 2--10 keV band corrected for absorption is estimated to be $1.5\times10^{40}$ erg s$^{-1}$. The error on the absorption column density introduces uncertainties in the correction of absorption, and the allowed range for the 2--10 keV luminosity is (1.2--2.9)$\times10^{40}$ erg s$^{-1}$. This luminosity is in the range for low-luminosity AGNs (e.g., \citealt{ho01,terashima02,cappi06}).
Thus the observed properties are in accordance with a low-luminosity AGN obscured by Compton-thin matter.

The luminosity of the X-ray source at the S nucleus is much higher than that of a single X-ray binary containing a stellar mass black hole or a neutron star.
If the X-ray source is a single compact object, its mass should be greater than 
\[
1000M_{\odot}
\left(
\frac{1.0}{
\lambda_{\rm Edd}
}
\right)
\left(
\frac{\kappa_{\rm 2-10~keV}}{10}
\right)
\left(
\frac{L_{\rm 2-10~keV}}{1.5\times10^{40} ~{\rm erg s}^{-1}
}
\right),
\]
where $\lambda_{\rm Edd}$, $\kappa_{\rm 2-10~keV}$, $L_{\rm 2-10~keV}$ are
the Eddington ratio, bolometric correction factor, and intrinsic luminosity in 2--10 keV, respectively.
Note that we assumed an Eddington luminosity of $1.5\times10^{38}$ erg
s$^{-1}$ for one solar mass, which is based on an assumption of a
Hydrogen-to-Helium ratio of 0.76:0.23 by weight. The bolometric correction
factor $\kappa_{2-10 \rm keV}$ is known to be dependent on a luminosity
and an Eddington ratio. Studies of AGN SEDs show that $\kappa_{2-10
\rm keV}$ is 10--20 for low-luminosity AGNs accreting at an Eddington
ratio of $<$0.1 \citep{vasudevan09,ho08}.

The X-ray fluxes of the S nucleus for the two observations with a 3 year interval are consistent with each other within errors. 
A systematic analysis of AGN X-ray light curves of a large sample shows that
a high percentage of low-luminosity AGN do not show significant variability
\citep{gm12}, and the absence of variability is 
consistent with
X-ray emission being from a low-luminosity AGN.
On the other hand, the absence of variability might be explained by the possibility that the X-ray emission comes from multiple sources.
If there are tens of stellar mass black holes well within the size of the PSF of {\it Chandra} (0\farcs 49 or 83~pc
at the distance of NGC~3256),
the luminosity measured could be explained. The very hard spectra observed, however, are not compatible with any spectral states
of stellar mass black hole binaries (\citealt{mcclintock06,done07}) 
and superposition of X-ray spectra suffering from different absorption column densities is required.
A large number of X-ray binaries are expected in starburst galaxies.
The integrated luminosity of high mass X-ray binaries, which have a hard
X-ray spectrum, formed by starburst activity is related to SFR \citep{ranalli03,grimm03,gilfanov04}.
The scaling law of \cite{grimm03} between SFR and integrated X-ray luminosity ($L_{\rm 2-10~keV}$) for low SFR values ($<4.5$ M$_{\odot}$ yr$^{-1}$) predicts $L_{\rm 2-10~keV}$ of $6\times10^{38}$ erg s$^{-1}$ for SFR of $\simeq 0.43$ M$_{\odot}$ yr$^{-1}$ (\S \ref{summary_MIR_SED_modeling}).
The observed luminosity is about a factor of $\sim 20$ larger than this expectation.
{\it Chandra} images of nearby starburst galaxies
indeed show many point sources (e.g., \citealt{griffiths00,strickland00,bauer01}). Such point sources, however, are distributed over a region of several hundreds of parsecs, which is much larger than the limit on the source size of the S nucleus. 
A single nuclear source is therefore more conceivable, 
although the possibility of superposition of multiple sources cannot be completely excluded.

A two component thermal plasma model also explains the shape of the observed spectra (\S \ref{thermal_model}). Thermal plasma emission with a temperature lower than $\sim 1$ keV is commonly observed in starburst galaxies (e.g., \citealt{strickland06}). Such a component is extended to the scale of host galaxy and intrinsic absorption is generally small. Our spectral fit indicates that emission from plasma with $kT<1 $ keV should be significantly absorbed by a column density of $1.3\times10^{22}$ cm$^{-2}$ or higher if such a component exists. Thus the observed properties are unusual to be interpreted as starburst galaxies in terms of the small source size and low-$kT$ plasma confined and hidden behind absorbing matter.

\subsubsection{Arguments at Other Wavelengths}

In addition to our MIR and X-ray evidence for the AGN, recent radio and NIR observations also suggest the AGN activity in the S nucleus.
\cite{sakamoto14} found a highly collimated bipolar jet-like outflow of molecular gas from the S nucleus using {\it ALMA}.
It extends up to 4\arcsec~(700~pc) from the nucleus and is associated with a bipolar spur of radio continuum emission.
They inferred from the morphology and kinematics of the molecular outflow and from the radio feature a jet-driving AGN in the S nucleus.
They also found that the continuum spectral slope at 860~$\mu$m is flatter at the S nucleus than at the N nucleus.
More synchrotron emission from an AGN in the S nucleus would explain the difference.
Similar bipolar outflow from the S nucleus was also detected in the NIR H$_2$ 1--0 $S$(1) line by \cite{emonts14}, and was suggested to be AGN-driven from the energetics and high-mass loading factor.

In addition to the X-ray to MIR SED (\citealt{lira02}; see also \S \ref{X_MIR} below), the X-ray to radio SED has been used to distinguish types of nuclear activities because these wavelengths are least sensitive to extinction and contamination by stellar activities is minimal there \citep{terashima03}.
\cite{neff03} found the ratio of 6~cm radio to 2--10 keV X-ray luminosities, $R_{\rm X}$ ($=L_{\rm R}/L_{\rm X}=\nu L_{\rm \nu}$ (5~GHz)/$L_{\rm 2-10~keV}$)$=1\times 10^{-2}$, for the compact source in the S nucleus from their {\it VLA} observations and the {\it Chandra} X-ray results of \cite{lira02}.
By comparing with the ratios of various kinds of Galactic and extragalactic sources, they argued that the compact source is most likely a low-luminosity AGN, although a possibility of a collection of supernova remnants cannot be rejected.
Our new X-ray luminosity is only 26\% larger than that of \cite{lira02}, and the revised $R_{\rm X}$ remains in the range of low-luminosity AGNs as \cite{neff03} discussed.

\subsection{Distribution of Dusty Material in and around the S Nucleus\label{dust_distribution}}

Our absorption magnitude and column density toward the AGN in the S nucleus, $A_{\rm V}\sim 80$ mag and $N_{\rm H} \sim10^{23}$ cm$^{-2}$, are much larger than previous estimates at NIR--MIR wavelengths.
They are $A_{\rm V} = 5.3$ and 10.7 mag from the $JHK'L'$ continuum and line photometries, respectively \citep{kotilainen96}, 16 mag from the {\it HST} NICMOS $H$--$K$ nuclear color measurement \citep{lira02}, and $15\pm 5$ mag from the {\it HST} NICMOS line flux ratio \citep{ah06a}.
Also, \cite{lira08} estimated $A_{\rm V} \sim 10$ mag on the basis of MIR SED fitting within 3\farcs 6--4\farcs 0 apertures.
It seems that the previous measurements of $A_{\rm V} = 5$--15 mag (i.e, $N_{\rm H} \sim 10^{22}$ cm$^{-2}$) correspond to the extended dust lanes over 1~kpc scale (e.g., \citealt{zepf99,moran99,lipari00,ah02,ah06a,sakamoto14}) or dusty circumnuclear star-forming region (\S \ref{MIR_SED_modeling}).
In our MIR SED analysis, we applied mild extinction corresponding to $A_{\rm V} \simeq 5$--8 mag on the starburst component of the AGN-starburst composite model (\S \ref{agn_starburst_composite_model}).
This component is most likely spatially extended because \cite{ah06b} reported a slightly extended structure in a 0\farcs 30 FWHM resolution (51~pc) image of the S nucleus at 8.74~$\mu$m.
We also found a slightly extended component, besides an unresolved core, in the NICMOS 2.2~$\mu$m image (0\farcs 75 FWHM or 128~pc) (\S \ref{infrared_morphology}).
In our X-ray analysis, we showed that the lightly absorbed component of the dual-component model suffers from $N_{\rm H,1}=0.34^{+1.56}_{-0.25}\times10^{22}$ cm$^{-2}$ (\S \ref{powerlaw_model}), matching the values from previous and our measurements at NIR--MIR.
Additional heavy extinction on the AGN is required to account for both the deep silicate absorption at the 0\farcs 36 scale and the heavy absorption on another component of the dual-component model in the X-ray analysis.
Such a heavily absorbed region is most likely within the central $\lesssim 0.5$\arcsec~of the S nucleus.
At MIR, the silicate absorption becomes much shallower at arcsec scale (\citealt{ds10}; \S \ref{MIR_SED_modeling}).
At X-ray, high spatial-resolution {\it Chandra} spectra show that the column density on a heavily absorbed component ($N_{\rm H,2}=7^{+19}_{-3}\times10^{22}$ cm$^{-2}$) is much larger than that on the power-law component in the whole-aperture {\it XMM-Newton} spectrum ($0.14\pm 0.03 \times10^{22}$ cm$^{-2}$; \citealt{ps11}).
For comparison, from high spatial-resolution {\it ALMA} molecular-gas imaging, \cite{sakamoto14} obtained a column density of
$N_{\rm H} = 10^{23}$--$10^{24}$ cm$^{-2}$
to the S nucleus in their 0\farcs 5 (80~pc) beam.

Both MIR and X-ray analyses strongly suggest that large amount of dusts distributes immediately around the AGN in a form of thick and smooth shell or torus.
ULIRGs generally show deeper silicate absorption ($S_{\rm 9.7\mu m} \ll -1$) than in AGNs (e.g., \citealt{hao07}), and such deep absorption requires that the energy source is deeply embedded in dust that is both optically and geometrically (along radial direction) thick, as well as geometrically smooth \citep{levenson07,nenkova08a,nenkova08b,sirocky08}.
For example, slab geometry cannot make the absorption $S_{\rm 9.7\mu m} < -1.1$ even with $\tau_{\rm V}=1000$, because one needs temperature gradient within the dusty region and the illuminated surface of the dust should be hidden from our direct views \citep{levenson07}.
Also, clumpy medium cannot make the absorption very deep because illuminated surface of individual clumps can be directly seen/illuminate dark surface of other clumps \citep{nenkova08a,nenkova08b}.
The observed upper limit of the silicate absorption depth at the 0\farcs 36 resolution, $S_{\rm 9.7\mu m} < -3$ \citep{ds10}, where the bottom of the feature was not determined due to very strong absorption and the sensitivity, is already in the range of the most heavily absorbed sources in the local universe \citep{hao07}.
Our MIR SED analysis implies a much deeper depth ($S_{\rm 9.7\mu m} = -9$ -- $-13$ where $S_{\rm 9.7\mu m} \equiv -\tau_{\rm 9.7}$ in our assumed geometry).
This strengthens a need for optically and geometrically thick dusty region immediately surrounding the AGN.
On the other hand, our X-ray spectrum analysis with a dual-component model suggests that a classical picture of type-2 AGNs with a dusty torus immediately surrounding a supermassive black hole (e.g., \citealt{antonucci93}) applies to the S nucleus.
Firstly, the elevated column density at the S nucleus is compatible with a typical dusty torus of Compton-thin type-2 AGNs 
(e.g., \citealt{awaki91,turner97,guainazzi05,fukazawa11}).
Secondly, the model also suggests that $\simeq 6$\% of the primary X-ray spectrum is seen unabsorbed by the heavy absorber, and this component can be interpreted as emission scattered by ionized medium in the opening part of the putative obscuring torus around the AGN.
Such phenomenon is commonly seen in typical Seyfert 2s \citep{turner97,noguchi10}. 
To reproduce deep silicate absorption with such a torus, the torus must be geometrically thick enough along the height direction, as well as along the radial direction, or our viewing angle must be close enough to the equatorial plane of the torus, so that the inner illuminated surface is hidden from our line-of-sight.

\subsection{Revisiting X-Ray-to-MIR SED\label{X_MIR}}

The S nucleus is known to be X-ray-quiet, and the fact has been considered to be evidence against an AGN there \citep{lira02}, although the galaxy as a whole is among the most luminous X-ray sources without confirmed AGN in the local universe \citep{moran99,lira02,ps11}.
\cite{lira02} revealed a compact X-ray source at the S nucleus with the first {\it Chandra} observation and reported the neutral Hydrogen column density obscuring the nucleus to be $N_{\rm H} = 5 \times 10^{22}$ cm$^{-2}$ (best fit)--$1 \times 10^{23}$ cm$^{-2}$ (acceptable).
By using a radio--FIR--MIR-X-ray SED of the S nucleus, \cite{lira02} showed that the observed X-ray luminosity is higher than that of typical starbursts but is at least two orders of magnitude lower than expected for a classical Seyfert nucleus.
\cite{alexander05} plotted the data of \cite{lira02} for the total aperture in the correlation diagram between IR and (absorption-corrected) X-ray luminosities together with well-studied local starbursts and AGNs.
They demonstrated that the X-ray luminosity of NGC~3256 between 0.5 keV and 8 keV is well below the range of AGNs but the galaxy is slightly more X-ray luminous than the starburst galaxies.

Our AGN-starburst composite model indicates that the observed SED is heavily contaminated at MIR by the circumnuclear star formation, and we found that the AGN component alone is consistent with typical AGNs in terms of X-Ray-to-MIR SED.
\cite{horst06,horst08} demonstrated that contamination by the circumnuclear starburst at MIR must be removed when evaluating AGN SEDs.
By using high spatial-resolution MIR photometries, \cite{horst06,horst08} and \cite{gandhi09} showed that the ratio of the MIR luminosity at 12.3~$\mu$m to the absorption-corrected X-ray luminosity at 2--10 keV scatters within a range of log $L_{\rm 12.3~\mu m}$/$L_{\rm 2-10~keV} = -0.5$ to 1.5, with a mean of $0.61 \pm 0.31$ among both type-1 and -2 Seyfert nuclei.
In the case of the S nucleus, we adopt the T-ReCS flux at 12.3~$\mu$m within the 0\farcs 36 aperture \citep{ds10}
and the absorption-corrected X-ray luminosity of the dual-component model (Model D; \S \ref{x_ray_evidence}).
Then we find log $L_{\rm 12.3~\mu m}$/$L_{\rm 2-10~keV} \sim 0.8$--1.2 ($=1.1$ with the best-fit 2--10 keV luminosity) for the AGN in the S nucleus.
It is in the range of ratios that \cite{horst08} and \cite{gandhi09} obtained for Seyfert nuclei.
Therefore, we conclude that the apparent X-ray quietness of the S nucleus should no longer be considered as evidence against AGN in the nucleus.

\subsection{Comparison with Previous Results}

Although our conclusion of the presence of an AGN in the S nucleus appears contradictory to most of the earlier studies, they are consistent with each other if we consider different aperture sizes and contamination from circumnuclear star formation.
\cite{lira02,lira08} noted the 4.5~$\mu$m excess at the S nucleus among the three IRAC channels.
\cite{lira02} also noted, by adding information of ground-based $JHK'L'N$-band photometry, that the S nucleus is in the 1--10~$\mu$m SED in $\nu F_{\rm \nu}$ flatter than typical starburst galaxies and is closer to AGNs.
However, they concluded that the AGN contribution is insignificant, if any.
This is because in \cite{lira02} their FIR--MIR--X-ray SED for the S nucleus is not consistent with the SEDs of typical Seyfert 2 nuclei, and in \cite{lira08} their T-ReCS $N$-band spectrum with a 1\farcs 3 slit shows evidence for star formation (e.g., PAH features).
In our work, we showed that the circumnuclear star formation dominates the MIR flux at a $\sim 3$\arcsec~scale, and it is not surprising that the T-ReCS spectrum from the 1\farcs 3 slit shows PAH features because the T-ReCS flux and the IRS flux within the 3\farcs 6 aperture are almost the same (\S \ref{MIR_SED_modeling}).
We also showed that the X-ray-to-MIR luminosity ratio is consistent with typical AGNs if we consider only the AGN contribution at MIR after excluding contribution from the circumnuclear star-forming regions (\S \ref{X_MIR}).

Our deeper {\it Chandra} spectrum enabled us to better constrain the X-ray characteristics owing to improved statistics.
On the basis of their first {\it Chandra} observation, \cite{lira02} presented spectral analysis of a composite X-ray spectrum of three hard sources including the S nucleus, and 
constrained the spectral shape of the S nucleus using X-ray hardness.
We used two {\it Chandra} data sets totaling $\simeq 2.7$ times more effective exposure time, and were able to obtain spectra of the S nucleus alone. The relatively flat spectra we obtained are 
qualitatively consistent with the results by \cite{lira02}. They assumed an absorbed power law model to constrain the allowed ranges of
the photon index and absorption column density. Our spectral fits indicate that a single-component absorbed power law model gives an extremely flat photon index ($\Gamma < 0.25$), otherwise the quality of fit becomes significantly worse. The absence of a strong Fe-K fluorescent line
in our spectrum strongly prefers
the model consisting of two components, one of which is absorbed by Compton-thin matter ($N_{\rm H} \sim 7\times10^{22}$ cm$^{-2}$), 
rather than a single absorbed power law.

The integrated spectrum of NGC~3256 measured with {\it XMM-Newton} provided a strong constraint on the flux of an Fe-K fluorescent line \citep{ps11}. Their upper limit on the flux is $3.5\times10^{-15}$ erg cm$^{-2}$ s$^{-1}$ or $3.4\times10^{-7}$ photons cm$^{-2}$ s$^{-1}$. Our limit on the Fe-K flux ($3.6\times 10^{-7}$ photons cm$^{-2}$ s$^{-1}$ for Model D) is almost the same as that obtained by \cite{ps11}. We obtained spectra of the S nucleus alone, in contrast to the integrated spectra previously reported,
and succeeded to set a limit on an Fe-K EW ($<190$ eV for Model C and $<550$ eV for Model D) that excludes the possibility of the presence of a Compton-thick AGN.

The AGN in the S nucleus is energetically much less important at MIR than the rest within NGC~3256 (\S \ref{summary_MIR_SED_modeling}), and previous estimates of the fractional AGN contribution within the galaxy seem not sensitive enough to detect the AGN.
\cite{ah12} made SED modeling of the IRS spectrum extracted over a kpc-scale region to explore possible AGN contribution.
They concluded that the AGN contributions to the whole-aperture MIR (at 6 and 24~$\mu$m) and IR luminosities are very small ($<5$\% and $<1$\%, respectively).
Our estimates of the AGN contribution to the S nucleus at $\sim 3$\arcsec~scale are $\sim 24$\% and $\sim 2$\% at 6 and 24~$\mu$m, respectively (\S \ref{summary_MIR_SED_modeling}).
Since the entire-galaxy IRS spectrum of \cite{ah12} is about 8 times brighter than our nuclear ones (\S \ref{irac_data_analysis}, \ref{irs_data_analysis}), the AGN contributes to $\sim 3$\% and $\sim 0.3$\% of the whole-aperture MIR luminosities at 6 and 24~$\mu$m, respectively.
Therefore, our results are consistent with those of \cite{ah12}.

The absorption corrected X-ray luminosity of the S nucleus for the dual absorber model with Seyfert-like power-law continua (Model D),
which we conclude most plausible, is $1.5\times10^{40}$ erg s$^{-1}$.
It is about 20\% of the total X-ray luminosity integrated over NGC 3256,
$7.4\times10^{40}$ erg s$^{-1}$, measured with {\it XMM-Newton}
\citep{ps11}.

\section{SUMMARY AND CONCLUSIONS}

NGC~3256 is the most luminous LIRG in the local universe ($z<0.01$), and it is a merging galaxy with two (N and S) nuclei.
Presence of an AGN in the S nucleus has been controversial.
We examined spectrophotometric characteristics of the S nucleus at both near--mid-infrared (NIR--MIR) and X-ray using archival and published data, and found several pieces of evidence to support a low-luminosity AGN obscured by Compton-thin matter.
The following are our findings and their implications.

We found in IRAC flux ratio maps that the S nucleus shows distinct photometric properties at 3.6--8.0~$\mu$m, most notably in its excess of 4.5~$\mu$m flux.
In contrast, the N nucleus is similar in colors to the star-forming regions within the host galaxy.
We applied the IRAC color-color diagrams for AGN diagnostics to the nuclear photometries for each nucleus, and found that the N and S nuclei show starburst- and AGN (power-law)-like SEDs, respectively.
This difference originates from the 4.5~$\mu$m excess at the S nucleus.

Using high-resolution {\it HST} NICMOS images, we extracted a compact source at the S nucleus by subtracting the stellar component at 2.2$~\mu$m.
The S nucleus consists of an unresolved (at 0\farcs 26 FWHM) and a resolved (0\farcs 75 FWHM or 128~pc after subtracting the instrumental resolution) component.
The corresponding structures are not seen at 1.6~$\mu$m and shorter wavelengths.
Because of its position and size, we identify its unresolved core with the compact MIR core (at 0\farcs 30 FWHM resolution at 8.74~$\mu$m).
The flux of the 2.2$~\mu$m core is consistent with our AGN spectral energy distribution (SED) model.

We analyzed the IRS nuclear spectrophotometries together with the IRAC nuclear photometries at 3.6--14.5~$\mu$m.
The IRS spectrum of the S nucleus shows bluer colors at $<6$~$\mu$m with respect to both the N nucleus and the IRS starburst template of \cite{brandl06} in a consistent way with the IRAC SED.
We conducted SED modeling of the S nucleus to reproduce the high spatial-resolution (0\farcs 36 aperture) T-ReCS $N$-band spectrophotometry \citep{ds10} as well as the nuclear IRAC 4.5~$\mu$m and IRS data.
Our AGN-starburst composite model (a heavily absorbed power-law AGN SED superposed on a mildly absorbed starburst-powered LIRG SED template) successfully reproduces both the deep silicate absorption at 9.7~$\mu$m within the 0\farcs 36 aperture and the distinct PAH features with the 4--6~$\mu$m excess at $\sim 3$\arcsec~scale.
We estimated $A_{\rm V}$ toward the AGN to be as large as 80 mag.
All the MIR results point toward a heavily absorbed (but Compton-thin) type-2 AGN in the S nucleus.
The AGN is most likely accompanied by circumnuclear star-forming regions.

We obtained a deep {\it Chandra} X-ray spectrum with $\simeq 2.7$ times more exposure time than in the previous study, and performed spectral analysis of a point-like source at the S nucleus.
We found that a dual-component (for primary and scattered ones) power-law model successfully fits the hard spectrum.
The inferred column density that is associated with the heavily absorbed primary component is $N_{\rm H}=7^{+19}_{-3} \times10^{22}$ cm$^{-2}$, which is in the range typically observed in Seyfert 2s and is consistent with our MIR measurement.
The lightly absorbed component can be interpreted as emission scattered by ionized medium in the opening part of the obscuring material around the AGN.
The successful dual-component model suggests a Compton-thin type-2 AGN in the S nucleus.
We also examined three other models for the S nucleus, namely a Compton-thick AGN, a single X-ray emitting compact source or a collection of such sources in the S nucleus, and thermal plasma with $kT<1$ keV.
We found that all of them are either rejected or less likely.
A Compton-thick AGN is rejected because of a limit on EW of a fluorescent Fe-K emission line at 6.4 keV ($<190$ eV for a reflection-dominated model).
Models with compact stellar source(s) are less likely because of the observed high luminosity, hard spectrum, and/or small region size well within the {\it Chandra} resolution (0\farcs 49 or 83~pc), although we cannot reject possibilities of models with multiple compact stellar sources.
A model with the thermal plasma seems unusual, although it is statistically possible, because the low-$kT$ plasma needs to be confined and hidden behind absorbing matter.

Our model with AGN is quantitatively consistent with most of the earlier studies that found no AGN in the S nucleus if we consider different aperture sizes and the star-forming circumnuclear region that dominates the MIR flux at $\sim 3$\arcsec~scale.
In our best MIR SED model (the AGN-starburst composite model), the AGN contributes to $\sim 2$\% and $\sim 0.2$\% of the whole-aperture MIR luminosities at 6 and 24~$\mu$m, respectively.
In our best X-ray model (a dual absorber model with Seyfert-like power-law continua), the absorption corrected X-ray luminosity of the S nucleus is about 20\% of the total X-ray luminosity integrated over NGC 3256.
In particular, we showed that the X-ray to MIR luminosity ratio of the AGN is consistent with that of typical Seyfert 2s if we exclude contribution of MIR flux from the circumnuclear star formation.

\acknowledgments

We thank our referee for useful comments and suggestions to improve this paper.
This work is supported by MOST grants 100-2112-M-001-001-MY3 (Y.O.) and 102-2119-M-001-011-MY3 (K.S.).
This work is based in part on observations made with the {\it Spitzer} Space Telescope, obtained from the NASA/IPAC Infrared Science Archive, both of which are operated by the Jet Propulsion Laboratory, California Institute of Technology under a contract with the National Aeronautics and Space Administration.
This research has made use of data obtained from the Chandra Data Archive and the Chandra Source Catalog, and software provided by the Chandra X-ray Center (CXC) in the application package CIAO.
Based on observations made with the NASA/ESA Hubble Space Telescope, and obtained from the Hubble Legacy Archive, which is a collaboration between the Space Telescope Science Institute (STScI/NASA), the Space Telescope European Coordinating Facility (ST-ECF/ESA) and the Canadian Astronomy Data Centre (CADC/NRC/CSA).

{\it Facilities:} \facility{Spitzer (IRAC)}, \facility{Spitzer (IRS)}, \facility{Chandra (ACIS)}, \facility{HST (NICMOS)}.

\clearpage

\clearpage

\begin{table}
\begin{center}
\caption{IRAC photometry results.\label{tab_irac_flux}}
\begin{tabular}{ccc}
\tableline\tableline
Channel & \multicolumn{2}{c}{Flux (mJy)\tablenotemark{a}} \\
 & N nucleus & S nucleus \\
\tableline
\multicolumn{3}{c}{Nuclear photometry\tablenotemark{b}} \\
\tableline
3.6~$\mu$m & $17.0 \pm 0.2$ & $9.4 \pm 0.1$ \\
4.5~$\mu$m & $17.9 \pm 0.2$ & $24.5 \pm 0.2$ \\
5.8~$\mu$m & $63.5 \pm 0.9$ & $41.0 \pm 0.6$ \\
8.0~$\mu$m & $203 \pm 22$\tablenotemark{c} & $51.0 \pm 2.7$ \\
\tableline
\multicolumn{3}{c}{Adjusted for the IRS spectra\tablenotemark{d}} \\
\tableline
3.6~$\mu$m & $34.7 \pm 0.4$ & $25.2 \pm 0.2$ \\
4.5~$\mu$m & $36.6 \pm 0.4$ & $41.2 \pm 0.3$ \\
5.8~$\mu$m & $130 \pm 1.9$ & $100 \pm 1.1$ \\
8.0~$\mu$m & $415 \pm 44$\tablenotemark{c} & $240 \pm 20$ \\
\tableline
\end{tabular}

\tablenotetext{a}{Uncertainties are for $\pm 1\sigma$.}
\tablenotetext{b}{Photometry with 2\farcs 8 diameter aperture with aperture correction.}
\tablenotetext{c}{The photometry is made on the image corrected for the saturation at the N nucleus.
See the main text for the details.}
\tablenotetext{d}{IRAC fluxes adjusted for the same level of circumnuclear flux contamination for the IRS resolution.
See the main text for the details of the adjustment.
}
\end{center}
\end{table}

\begin{table}
\begin{center}
\caption{Results of the double-Gaussian model fitting to the S nucleus on the stellar-component subtracted F222M image.\label{tab_nicmos_measurement}}
\begin{tabular}{ccc}
\tableline\tableline
 & compact source & extended source \\
\tableline
Position & \multicolumn{2}{c}{(J2000)} \\
RA    & \multicolumn{2}{c}{10:27:51.24\tablenotemark{a}} \\
DEC & \multicolumn{2}{c}{-43:54:19.2\tablenotemark{a}} \\
source size & \multicolumn{2}{c}{(FWHM in arcsec)} \\
                   & unresolved\tablenotemark{b} & $0\farcs 75 \pm 0 \farcs 01$ \tablenotemark{c}\\
Flux & \multicolumn{2}{c}{(mJy)\tablenotemark{d}} \\
        & $0.65\pm 0.02$ & $5.33 \pm 0.16$ \\
\tableline
\tablenotetext{a}{Uncertainties (0\farcs 2) are dominated by global astrometry solution. See the main text for the details.}
\tablenotetext{b}{Assumed to be unresolved at 0\farcs 26 FWHM.}
\tablenotetext{c}{After correcting for the instrumental resolution of 0\farcs 26 FWHM.}
\tablenotetext{d}{Uncertainties are for $\pm 1\sigma$.}
\end{tabular}
\end{center}
\end{table}

\begin{table}
\begin{center}
\caption{MIR spectrophotometry model fitting results.\label{tab_sed_fitting_results}}
\begin{tabular}{cccccccc}
\tableline\tableline
Extinction curve & Data to fit & models\tablenotemark{a} & Single-SB\tablenotemark{b} & \multicolumn{2}{c}{AGN+SB} & \multicolumn{2}{c}{SB+SB} \\
 & & & & 0\farcs 36 aperture & $\simeq 3$\arcsec~aperture & 0\farcs 36 aperture & $\simeq 3$\arcsec~aperture \\
\tableline
\cite{ct06}\tablenotemark{c} & 4.5~$\mu$m+IRS+T-ReCS & $\tau_{\rm 9.7}$ & $1.04\pm 0.01$ & $9.4\pm 0.4$ & $0.68\pm 0.01$ & $6.8\pm 0.3$ & $0.98\pm 0.02$ \\
                                             &                                  & reduced $\chi^2$ & 4.09 & 1.4 & 1.4 & 1.2 & 4.4 \\
                                             & IRS+T-ReCS            & reduced $\chi^2$ & 0.90 & 1.4 & 0.77 & 1.2 & 0.87 \\
Original PAHFIT                   & 4.5~$\mu$m+IRS+T-ReCS & $\tau_{\rm 9.7}$ & $0.00$ & $8.0\pm 0.4$ & 0.0 & $6.0\pm 0.3$ & 0.0 \\
                                             &                                  & reduced $\chi^2$ & 5.80 & 3.6 & 5.2 & 2.7 & 6.0 \\
                                             & IRS+T-ReCS            & reduced $\chi^2$ & 1.09 & 3.6 & 1.08 & 2.7 & 1.11 \\
Modified PAHFIT\tablenotemark{d} & 4.5~$\mu$m+IRS+T-ReCS & $\tau_{\rm 9.7}$ & $1.21\pm 0.05$ & $12.7\pm 0.5$ & $0.48\pm 0.04$ & $8.3\pm 0.4$ & $1.10\pm 0.02$ \\
                                             &                                 & reduced $\chi^2$ & 3.58 & 1.2 & 0.79 & 1.2 & 3.5 \\
                                             & IRS+T-ReCS           & reduced $\chi^2$ & 0.90 & 1.2 & 0.74 & 1.2 & 0.86 \\
\tableline
\tablenotetext{a}{Models: Single-SB = absorbed single-starburst model; AGN+SB = AGN-starburst composite model; SB+SB = starburst-starburst composite model.}
\tablenotetext{b}{T-ReCS data are not used for this model.}
\tablenotetext{c}{Toward Galactic center.}
\tablenotetext{d}{Modified to reproduce the extinction curve of \cite{fritz11}. See the main text for the details.}
\end{tabular}
\end{center}
\end{table}

\begin{table}
\begin{center}
\caption{{\it Chandra} Observation Log.\label{tab_chandra_log}}
\begin{tabular}{cccc}
\tableline\tableline
Observation ID\tablenotemark{a} & Date\tablenotemark{b} & Exposure\tablenotemark{c} & Counts\tablenotemark{d} \\
	&	& (ksec) & \\
\tableline
835 & 2000 Jan 5 & 16.2 & 39\\
3569 & 2003 May 23 & 27.2 & 62\\
\tableline
\tablenotetext{a}{Observation identification number.}
\tablenotetext{b}{Observation start date.}
\tablenotetext{c}{Exposure time after data screening.}
\tablenotetext{d}{Number of counts after background subtraction in 0.5-7 keV.}

\end{tabular}
\end{center}
\end{table}

\begin{deluxetable}{ccccccccccc}
\tablewidth{0pc}
\tablecaption{Results of X-ray spectral fits with power law model.\label{tab_chandra_pl_fit_results}}
\tabletypesize{\small}
\tablehead{
Model$^a$ & $N_{\rm H, 1}$ &$N_{\rm H, 2}$ & $f$ & Photon index & Norm. & Constant & $C$/dof & EW & $F_{\rm 2-10~keV}$ & $L_{\rm 2-10~keV}$ \\
	&	($10^{22}$ cm$^{-2}$) & ($10^{22}$ cm$^{-2}$) & & & & & & (eV)\\
(1) & (2) & (3) & (4) & (5) & (6) & (7) & (8) & (9)& (10) & (11)
}
\startdata
A & 0.054 ($<$0.97) & ... & ... & $-0.42^{+0.67}_{-0.39}$ & 0.067 & $0.94^{+0.40}_{-0.27}$ & 76.8/90 & $<$150 & 11.5 & 1.7 \\
B & 3.6 & ... & ... & 1.8$^a$ &2.15 & 0.92 & 94.5/91$^b$ & ... & ... & ... \\
C & 0 ($<0.27$) & ... & ... & 1.8$^a$ & 46.4 & $0.93^{+0.39}_{-0.27}$ & 78.1/91 & $<$190 & 9.26 & 1.4\\
D & $0.34^{+1.56}_{-0.25}$ & $7^{+19}_{-3}$ & $0.943^{+0.037}_{-0.149}$ & 1.8$^a$ & 2.93 & $0.95^{+0.40}_{-0.27}$ &78.6/89 & $<550$ & 6.33 & 1.5 \\
\enddata
\tablecomments{
(1) Spectral models.
The Galactic absorption ($N_{\rm H} = 9.14\times10^{20}$ cm$^{-2}$) is applied to all the models. 
(2), (3) Intrinsic absorption column density.
(4) The fraction of continuum absorbed by a large column density ($N_{\rm H,2}$).
(5) Photon index of power law component.
(6) Normalization of power law in units of $10^{-5}$ photons cm$^{-2}$ s$^{-1}$ at 1 keV. 
(7) Constant factor relative to the first observation.
(8) $C$ statistic and degrees of freedom.
(9) Upper limit on the equivalent width of fluorescent Fe-K$\alpha$ line in units of eV.
(10) Observed flux in the 2--10 keV band for the first observation in units of $10^{-14}$ erg s$^{-1}$ cm$^{-2}$.
(11) Luminosity corrected for absorption in the 2--10 keV band for the first observation in units of $10^{40}$ erg s$^{-1}$.
}
\tablenotetext{a}{Fixed parameter.
}
\tablenotetext{b}{Errors, limit on Fe-K line equivalent width, flux, and luminosity are not shown for this model since the value of $C$ statistic is significantly worse than those for other models.
}
\end{deluxetable}

\begin{deluxetable}{ccccccccccc}
\tablenum{6}
\tablewidth{0pc}
\tablecaption{Results of X-ray spectral fits with thermal model.\label{tab_chandra_thermal_fit_results}}
\tabletypesize{\small}
\tablehead{
Model$^a$ & $N_{\rm H}$ &$N_{\rm H}$ & $kT$ & Norm. & $kT$ & Norm.  & Constant & $C$/dof & $F_{\rm 2-10~keV}$ & $L_{\rm 2-10~keV}$ \\
	&	($10^{22}$ cm$^{-2}$) & ($10^{22}$ cm$^{-2}$) & & & & & & \\
(1) & (2) & (3) & (4) & (5) & (6) & (7) & (8) & (9)& (10) & (11)
}
\startdata
E & $5.1^{+3.3}_{-1.9}$ & ... & 64$^a$ ($>$9.4) & 6.71 & 64$^a$ ($>$1.2) & 0.428 & $0.93^{+0.39}_{-0.26}$ & 77.6/88 &6.98 & 1.4\\
F & $4.3^{+4.0}_{-2.0}$ & $1.3^{+1.0}_{-0.5}$    & 64$^a$ ($>$9.7) & 7.44  & 1.0$^b$  & 0.798 & $0.95^{+0.40}_{-0.28}$ & 77.3/88 & 7.06 & 1.4\\
\enddata
\tablecomments{
(1) Spectral models.
The Galactic absorption ($N_{\rm H} = 9.14\times10^{20}$ cm$^{-2}$) is applied to all the models. 
(2) Intrinsic absorption column density for heavily absorbed APEC component.
(3) Intrinsic absorption column density common to both APEC components.
(4) Temperature of heavily absorbed APEC component.
(5) Normalization of heavily absorbed APEC component in units of 
 $10^{-19}/(4\pi(D_{\rm A}(1+z))^2 / \int n_{\rm e} n_{\rm H} dV$,
 where $D_{\rm A}$ is the angular size distance to the source (cm), 
 $n_{\rm e}$ is the electron density (cm$^{-3}$), and $n_{\rm H}$ is the hydrogen density (cm$^{-3}$).
(6) Temperature of lightly absorbed APEC component.
(7) Normalization of lightly absorbed APEC component in units of 
$10^{-19}/(4\pi(D_{\rm A}(1+z))^2 / \int n_{\rm e} n_{\rm H} dV$.
(8) Constant factor.
(9) $C$ statistic and degrees of freedom.
(10) Observed flux in the 2--10 keV band for the first observation in units of $10^{-14}$ erg s$^{-1}$ cm$^{-2}$.
(11) Luminosity corrected for absorption in the 2--10 keV band for the first observation in units of $10^{40}$ erg s$^{-1}$.
}
\tablenotetext{a}{Pegged at the highest temperature allowed in the fit.
}
\tablenotetext{b}{Fixed parameter.}
\end{deluxetable}

\end{document}